\documentclass[journal]{IEEEtran}

\usepackage[T1]{fontenc}
\usepackage[table,xcdraw]{xcolor}
\usepackage{listings}
\usepackage{colortbl}
\usepackage{xurl}
\usepackage{fancyvrb}

\usepackage{nicefrac}
\usepackage{siunitx}
\usepackage{array,framed}
\usepackage{booktabs}
\usepackage{tabularx}

\usepackage{float, epsfig, wrapfig, graphicx, subcaption}
\usepackage{tikz}
\usepackage{pifont}

\definecolor {red}{RGB}{210, 80, 80}
\definecolor{green}{RGB}{80, 180, 80}

\usepackage{rotating}
\usepackage{ulem}

\newcommand{\emarkcircle}{
  \tikz[baseline=-0.5ex]\node[draw=green!60!black, fill=green!0, circle, inner sep=1pt]{\scalebox{0.9}{$\checkmark$}};
}

\newcommand{\cmarkcircle}{
  \tikz[baseline=-0.5ex]\node[draw=green!60!black, fill=green!30, circle, inner sep=1pt]{\scalebox{0.9}{$\checkmark$}};
}

\usepackage{textcomp,amssymb}
\usepackage{fix-cm}
\usepackage{latexsym,fancyhdr,url}
\usepackage{enumerate}
\usepackage{graphics}
\usepackage{xparse} % argument parsing -- \edist
\usepackage{multirow}
\usepackage{csvsimple}
\usepackage{balance}
\usepackage{amsmath, amssymb}
\usepackage{graphicx}
\usepackage{textcomp}

\usepackage{booktabs}
\usepackage{soul}
\usepackage{xspace}
\usepackage{mathtools}
\usepackage{wasysym}
\usepackage{algorithm}
\usepackage[noend]{algpseudocode}
\usepackage{kotex}
\usepackage{scalerel,stackengine}
\usepackage{makecell}
\usepackage{fontawesome}

\newcommand{\ourtool}{\texttt{AutoPatch}\xspace}

\usepackage{
  pgfplots,
  pgfplotstable
}

\usetikzlibrary{
  shapes.geometric,
  arrows,
  external,
  pgfplots.groupplots,
  matrix
}

\pgfplotsset{compat=1.9}
\usepackage{mathtools}

\DeclareMathAlphabet{\mathcal}{OMS}{cmsy}{m}{n}
\DeclareGraphicsExtensions{%
    .png,.PNG,%
    .pdf,.PDF,%
    .jpg,.mps,.jpeg,.jbig2,.jb2,.JPG,.JPEG,.JBIG2,.JB2}

\ifCLASSINFOpdf
\else
\fi
\begin{document}
\title{\texttt{AutoPatch}: Multi-Agent Framework for Patching Real-World CVEs Generated by Outdated LLMs}

\author{Minjae Seo$^\ast$,
        Wonwoo Choi$^\ast$,
        Seungwon Shin,
        Myoungsung You% <-this % stops a space
\thanks{$^\ast$ Minjae Seo and Wonwoo Choi contributed equally to this work.}%
\thanks{M. Seo is with Electronics and Telecommunications Research Institute.}% <-this % stops a space
% \thanks{M. Seo is with the Department
% of Electrical and Computer Engineering, Georgia Institute of Technology, Atlanta,
% GA, 30332 USA e-mail: (see http://www.michaelshell.org/contact.html).}% <-this % stops a 
\thanks{W. Choi is with Agency for Defense Development.}% <-this % stops a space
\thanks{S. Shin is with the School of Electrical Engineering, Korea Advanced Institute of Science and Technology.}%
\thanks{M. You is with the School of Electrical and Computer Engineering, University of Seoul. E-mail: famous@uos.ac.kr}%
% \thanks{Manuscript received April 19, 2005; revised August 26, 2015.}
}

\maketitle

\begin{abstract}
Large Language Models (LLMs) have emerged as promising tools in software development, enabling automated code generation and analysis.
However, their knowledge is limited to a fixed cutoff date, making them prone to generating code vulnerable to newly disclosed CVEs.
Frequent fine-tuning LLMs with newly disclosed CVEs is costly, and existing LLM-based approaches typically rely on oversimplified CWE examples and require providing explicit bug locations to LLMs, making them ill-suited for instantly patching real-world vulnerabilities in LLM-generated code.
To address these limitations, we propose \ourtool{}, a multi-agent framework designed to patch vulnerable LLM-generated code, particularly those introduced after the LLMs’ knowledge cutoff.
\ourtool{} integrates Retrieval-Augmented Generation (RAG) with a structured database of recently disclosed vulnerabilities, comprising 525 code snippets derived from 75 high-severity CVEs across real-world systems such as the Linux kernel, Chrome, and others.
\ourtool{} combines semantic and data flow analysis to identify the most relevant CVE and leverages enhanced Chain-of-Thought (CoT) reasoning to construct enriched prompts for verification and patching.
Our unified similarity model, which selects the most relevant CVE, achieves 91.8\% accuracy in CVE matching. \ourtool{} attains an F1-score of 90.3\% for vulnerability verification and an accuracy of 94.1\% in patching, while being over 50× more cost-efficient than traditional fine-tuning approaches.

\end{abstract}


% \fi
\IEEEpeerreviewmaketitle

\section{Introduction}

% \begin{figure}
%     \centering
%     \includegraphics[width=0.8\linewidth]{figures/Intro_v6.pdf}
%     \caption{The overall workflow of \ourtool{}.}
%     \label{fig:tot_intro}
%     \vspace{-0.1in}
% \end{figure}

\IEEEPARstart{L}{arge} Language Models (LLMs) have become integral tools in software development, exhibiting strong capabilities in automated code generation and debugging. Code generation LLMs, such as ChatGPT~\cite{openai_platform}, Codex~\cite{copilot_platform}, CodeLlama~\cite{roziere2023code}, and DeepSeek~\cite{guo2025deepseek}, are now widely adopted by developers. Consequently, over one million programmers actively used GitHub Copilot by 2024~\cite{hossen2024assessing}, demonstrating the substantial impact of these models on the software development lifecycle.

% Prominent examples include ChatGPT~\cite{openai_platform}, Codex~\cite{copilot_platform}, CodeLlama~\cite{roziere2023code}, \wwc{StarCoder2~\cite{li2023starcoder},} and DeepSeek~\cite{guo2025deepseek}. These models have been widely adopted by developers; Over one million programmers have actively adopted GitHub Copilot by 2024~\cite{hossen2024assessing}. This widespread adoption shows the profound impact of LLM-assisted programming, significantly accelerating the software development lifecycle.

% In parallel with the rise of LLM-assisted coding, the prevalence of software vulnerabilities has surged to unprecedented levels. According to NVD statistics, over 40,000 publicly disclosed vulnerabilities (CVEs) were reported in 2024 alone, representing a 38\% increase from the 28,818 reported in 2023~\cite{NVD_STAT_2024,CVE2024}. This trend continues into 2025, with 1,148 Linux kernel vulnerabilities and 38 critical Chrome flaws in just the first two months~\cite{CVE_Chrome_2025}

% for example, in just the first two months of 2025 (January and February), a total of 1,148 new Linux kernel vulnerabilities were recorded~\cite{Linux_2025}, and around the same time, Google Chrome users were confronted with 38 critical security flaws disclosed across multiple updates~\cite{CVE_Chrome_2025}.

While LLMs significantly accelerate software development, the prevalence of software vulnerabilities has concurrently risen at an unprecedented rate. In 2024 alone, over 40,000 publicly disclosed vulnerabilities were reported~\cite{NVD_STAT_2024}, and within just the first two months of 2025, 1,148 Linux kernel vulnerabilities and 39 critical Chrome flaws were disclosed~\cite{CVE_Chrome_2025}. Despite these trends, LLMs do not automatically learn about vulnerabilities discovered after their knowledge cutoff, the point beyond which no additional data is incorporated into training. Consequently, they may unwittingly suggest code that contains known security vulnerabilities because those issues were not part of their training data. Prior studies report that roughly 30\% of LLM-generated code suggestions include previously documented vulnerabilities~\cite{hossen2024assessing, yu2024fight}. Our analysis further confirms that even state-of-the-art LLMs can reproduce vulnerable patterns introduced after their cutoff date, as these vulnerabilities are absent from their training corpus (see Section~\ref{sec:motivation}). Consequently, without proper secure coding practices, naive reliance on LLMs can introduce outdated or insecure code, leading to severe security risks such as financial loss, service disruption, and data breaches~\cite{LLM_INSECURE}.

\begin{figure}[t]
    \centering
    \includegraphics[width=1.0\linewidth]{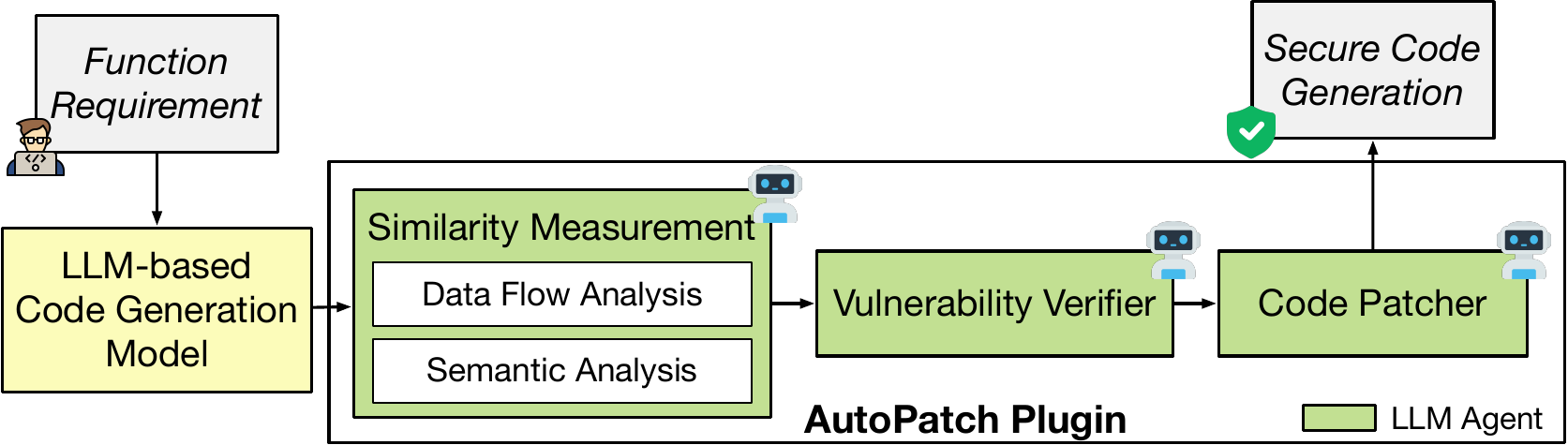}
    \caption{The overall workflow of \ourtool{}.}
    \label{fig:tot_intro}
    \vspace{-4mm}
\end{figure}

To address these concerns, one potential solution is to frequently fine-tune LLMs with newly disclosed vulnerability data, such as recent CVEs and their corresponding patches. However, this approach is prohibitively time-consuming and resource-intensive~\cite{nong2024automated,sheng2025llms,keltek2024lsast}, as LLM training or fine-tuning requires substantial GPU resources for extended periods. As an alternative, recent studies have explored prompt engineering techniques (e.g., Chain-of-Thought (CoT) prompting), which convey new CVE information through prompts without updating model parameters~\cite{nong2024chain,nong2024automated,yin2024thinkrepair,zhang2025patch}. Despite their promise, existing methods exhibit three major limitations. First, prior work~\cite{nong2024chain,nong2024automated} relies on simplified vulnerability examples representing only narrow CWE categories, which limits generalizability to real-world vulnerabilities and reduces effectiveness when applied to LLM-generated code. Second, many of these approaches require explicit bug locations as input rather than performing autonomous vulnerability detection~\cite{nong2024automated,zhang2025patch}, resulting in a human-in-the-loop workflow that demands extensive manual effort. Third, they place the full responsibility for vulnerability identification on LLMs alone, which constraints applicability in complex, real-world security environments~\cite{lyu2023faithful,ullah2024llms}. Consequently, these approaches predominantly reflect the perspective of a bug-testing workflow and overlook the needs of software developers, who require more practical and scalable mechanisms for vulnerability detection in real-world development settings.

To overcome these limitations, we propose \ourtool{}, a multi-agent-based system designed to identify real-world vulnerabilities in LLM-generated code and automatically apply secure patches, even for vulnerabilities disclosed after the model's training cutoff.
As illustrated in Fig.~\ref{fig:tot_intro}, \ourtool{} is structured as a security plugin for LLM-integrated IDEs and consists of three specialized LLM agents: the Similarity Analyzer, the Vulnerability Verifier, and the Code Patcher. When a developer provides a functional requirement, the LLM-integrated IDE generates an initial code snippet, which \ourtool{} subsequently analyzes using a structured, multi-stage workflow. To proactively detect vulnerabilities, the Similarity Analyzer agent first extracts key terms and contextual descriptions from the LLM-generated code and conducts semantic analysis by comparing these features against the semantic representations of recently disclosed vulnerabilities stored in a RAG DB, thereby calculating a \textit{semantic similarity} score.
% by comparing them against a RAG database (RAG DB) of recently disclosed vulnerabilities. 
In parallel, the agent performs data flow analysis on the LLM-generated code to enhance the understanding of the internal components' operations. It summarizes the flow of variables and function calls into symbolic representations that omit explicit naming, and calculates a \textit{data flow similarity} score by matching against the database.
% The derived representations are matched against the database to calculate the taint similarity score. 
The \textit{semantic similarity} score and the \textit{data flow similarity} score are then combined into a \textit{unified similarity} score. To optimize this process, we train a machine-learning model (unified model) which learns optimal weights via pairwise ranking loss, ensuring that relevant CVEs are consistently ranked above irrelevant ones.

% \mj{We maintain a RAG-enhanced database (RAG DB) storing structured information for each vulnerability, including the CVE ID, vulnerability type, original/patched code, and semantic/taint representations. In addition, it includes vulnerability-related variables and functions (with both raw text and vector representations), root cause explanations, and CoT prompts for both vulnerability detection and patching, enabling context-aware retrieval and reasoning throughout the pipeline.} 
%We maintain a RAG database storing structured information for each vulnerability to support context-aware verification and patching. Each entry includes the CVE ID, vulnerability type, and vulnerability-related variables and functions—shared components that provide a consistent semantic foundation across both tasks. For verification, entries include the original code and a verification CoT used to construct one-shot examples. For patching, they include the patched code and a corresponding patching CoT prompt. Upon identifying a match, the Vulnerability Verifier retrieves the corresponding entry and constructs a one-shot verification example to explain how the matched vulnerability manifests and its root cause, enriching the LLM query for more accurate assessment. If the generated code is deemed vulnerable, the Code Patcher constructs a corresponding one-shot patching example and queries the LLM to generate a secure revision. The revised code is then re-evaluated by the Vulnerability Verifier, and this cycle repeats until the code is verified to be free of vulnerabilities.

Upon identifying a match, the Vulnerability Verifier retrieves the corresponding entry from the RAG DB and constructs a one-shot verification example to explain how the matched vulnerability manifests and its root cause, enriching the LLM query for more accurate assessment. If the generated code is deemed vulnerable, the Code Patcher constructs a corresponding one-shot patching example and queries the LLM to generate a secure revision. The revised code is then re-evaluated by the Vulnerability Verifier, and this cycle repeats until the code is verified to be free of vulnerabilities.

% We implement a full prototype of the \ourtool{}. The unified similarity model is trained using the Adam optimizer, and multi-agent coordination with RAG-enhanced retrieval is achieved using LangChain and a PostgreSQL vector database. We evaluate with GPT-4o, Code Llama, DeepSeek, and o3-mini, generating a total of 525 code snippets across 75 recent high-severity CVEs (including Linux kernel, Chrome, and others), for which we manually construct a benchmark to ensure reliable assessment. GPT-4o achieves 89.3\% fidelity in recreating vulnerabilities; our similarity model achieves 91.8\% accuracy in matching CVEs. During verification, \ourtool{} with GPT-4o reaches an F1-score of 90.3\% for vulnerability detection. For patching, \ourtool{} with GPT-4o successfully patches vulnerable code with an accuracy of 94.1\%. Notably, fine-tuning at intervals of five CVEs across the entire dataset incurs a cost that is 5,230\% higher, demonstrating the substantial efficiency gains provided by our lightweight, plugin-based approach. 

\noindent\textbf{Contributions.} We make the following contributions:
\begin{itemize}

    \item We show, through a real-world example, that even state-of-the-art LLMs often generate insecure code with known vulnerabilities disclosed after their knowledge cut-off.
    
    \item We design and implement \ourtool{}, a novel multi-agent-based security framework capable of identifying known vulnerabilities in LLM-generated code and generating corresponding patches by leveraging RAG DB-assisted prompting and a unified detection model.
    
    \item We construct a new benchmark dataset for LLM-based vulnerability detection and patching, comprising 525 code snippets derived from 75 recent high-severity CVEs.
    
    \item We evaluate \ourtool{} using GPT-4o, Code Llama, DeepSeek, and o3-mini. Results show that \ourtool{} with GPT-4o achieves an F1-score of 90.3\% in vulnerability detection and a patching accuracy of 94.1\%. In addition, compared to fine-tuning with the entire dataset, \ourtool{} achieves a 5,230\% lower cost.
    \vspace{0.2in}

    % \item We propose \ourtool{}, a cost-efficient multi-agent framework that eliminates the need for fine-tuning of LLMs to handle newly disclosed CVEs.
    % \item We enhance vulnerability detection and patch generation accuracy by leveraging a high-severity CVE RAG database enriched with semantic and data flow analysis to identify relevant vulnerabilities, combined with advanced reasoning for verification and patch synthesis.

    % \item We implement a full prototype of \ourtool{} utilizing LangChain and a PostgreSQL vector database, and evaluate its effectiveness on recently disclosed high-severity CVEs
    % collected from real-world codebases.
\end{itemize}

\noindent\textbf{Organization.}
Section~\ref{sec:motivation} provides background on LLM-based code generation and introduces our motivating example. Section~\ref{sec:design} describes the overall design and workflow of \ourtool{}. Section~\ref{sec:Implementation} and Section~\ref{sec:evaluation} describe the implementation details and evaluation results, respectively. Section~\ref{sec:discussion} discusses the current limitations and future directions. Section~\ref{sec:related_work} reviews related work. Finally, Section~\ref{sec:conclusion} concludes this paper.

% To evaluate LLM-generated vulnerable code, we use Code Llama, DeepSeek Coder, DeepSeek-R1, GPT-4o, and OpenAI o3-mini, generating a total of 525 code snippets across 75 high-severity, recently disclosed vulnerabilities. Among them, GPT-4o achieves the highest fidelity, reimplementing vulnerable code with 89.3\% accuracy. Our unified model achieves 90.4\% accuracy in identifying real-world vulnerabilities. During verification, \ourtool{} integrated with GPT-4o achieves the highest F1-scores of 91.6\% for vulnerability-only detection and 89.5\% when jointly verifying vulnerability and CoT reasoning. Furthermore, \ourtool{} with GPT-4o successfully patches the vulnerable code with 95.0\% accuracy. Notably, fine-tuning at an interval of 5 CVEs across the entire set incurs a cost that is 5,230\% higher, demonstrating the substantial efficiency gains provided by our plugin-based approach.
% \vspace{0.25cm}

\section{Background and Motivation}

\begin{table}[t]
\centering
    \begin{minipage}{\linewidth}
    % \footnotesize
    \centering
    \scriptsize
    \caption{Knowledge cutoff of code generation models}
    \renewcommand{\arraystretch}{1.0} % Adjust row spacing slightly for readability
    \begin{tabular}{>{\centering\arraybackslash}m{2cm} 
                    >{\centering\arraybackslash}m{3cm} 
                    >{\centering\arraybackslash}m{2.5cm}} 
        \toprule
        \textbf{Base Model} & \textbf{Model Variant} & \textbf{Knowledge Cutoff} \\ 
        \midrule
        \multirow{4}{*}
        {ChatGPT~\cite{openai_platform}} 
        & o3-mini & Oct 2023 \\ 
        \cline{2-3}
        & \vspace{0.5ex} GPT-4o & \vspace{0.5ex}Oct 2023 \\ 
        \cline{2-3}
        & \vspace{0.5ex}GPT-4o Realtime & \vspace{0.5ex}Oct 2023 \\ 
        \cline{2-3}
        & \vspace{0.9ex}GPT-4 Turbo & \vspace{0.9ex}Dec 2023 \\ 
        \midrule
        \multirow{2}{*}{Codex~\cite{copilot_platform}} & Copilot (GPT-3.5 Turbo) & Sep 2021 \\ 
        \cline{2-3}
        & \vspace{1.0ex}Copilot (GPT-4o mini) & \vspace{1.0ex}Oct 2023 \\ 
        \midrule
        \multirow{2}{*}
        {Llama 3~\cite{llama_platform}} & Llama-3-8B & Mar 2023 \\
        \cline{2-3}
        & \vspace{0.9ex}Llama-3-70B & \vspace{0.9ex}Dec 2023 \\
        \midrule
        \multirow{1}{*}{DeepSeek~\cite{zhu2024deepseek}} & DeepSeek-Coder-V2 & Nov 2023 \\
        \bottomrule
    \end{tabular}
    
    \label{tab:codegen_models}
    \end{minipage}
\end{table}

\subsection{Code Generation Model}
Code generation models are a specialized subset of large language models (LLMs) designed to produce executable code from natural language descriptions. These models are trained on extensive datasets, including numerous open-source software repositories, enabling them to generate individual functions as well as complex multi-file programs. Notable examples include ChatGPT~\cite{openai_platform}, Codex~\cite{copilot_platform}, LLaMA~\cite{llama_platform}, and DeepSeek~\cite{zhu2024deepseek}. Many of these models are integrated into modern software IDEs (e.g., Visual Studio Code), allowing developers to interact with them directly. For instance, developers can request a new function by supplying code snippets and natural language specifications (or code comments), and the IDE inserts the generated function into the current project. By 2024, over one million developers had adopted GitHub Copilot as part of their workflow~\cite{hossen2024assessing}, substantially reducing the time and effort required for software development.

\subsection{Retrieval-Augmented Generation (RAG)}
Retrieval-Augmented Generation (RAG)~\cite{lewis2020retrieval} offers a compelling alternative to the resource-intensive fine-tuning tasks by enhancing pre-trained models with external, domain-specific data. Instead of modifying the model’s internal parameters, RAG integrates a retrieval mechanism that accesses up-to-date and relevant information from external sources during inference, thereby enriching the model's responses for specialized tasks. This minimizes computational demands and ensures that the model remains adaptive and context-aware, making it an attractive solution for applications requiring continuous updates and precision in domain-specific outputs.

\subsection{Motivation}
\label{sec:motivation}
% \begin{figure*}[t]
%     % \centering
%     % \hspace*{-0.1\linewidth} % adjust this value as needed
%     \includegraphics[width=1.0\textwidth]{figures/Motivation_1.pdf}
    
%     \caption{GPT-4o-2024-11-20 generated vulnerable code for CVE-2025-21671, exposing a dangling pointer that may cause a use-after-free vulnerability.}
%     \label{fig:motiv}
    
% \end{figure*}

% \begin{figure*}[t]
%     % \centering
%     % \hspace*{-0.1\linewidth} % adjust this value as needed
%     \includegraphics[width=1.0\textwidth]{figures/Motivation_2_v5.pdf}
    
%     \caption{GPT-4o-2024-11-20 generated secure code for CVE-2022-0185, correctly addressing the unsigned integer overflow that caused the vulnerability.}
%     \label{fig:motiv_2}
% \end{figure*}

\begin{figure*}[t]
    \centering
    % \hspace*{-0.1\linewidth} % adjust this value as needed
    \includegraphics[width=0.85\textwidth]{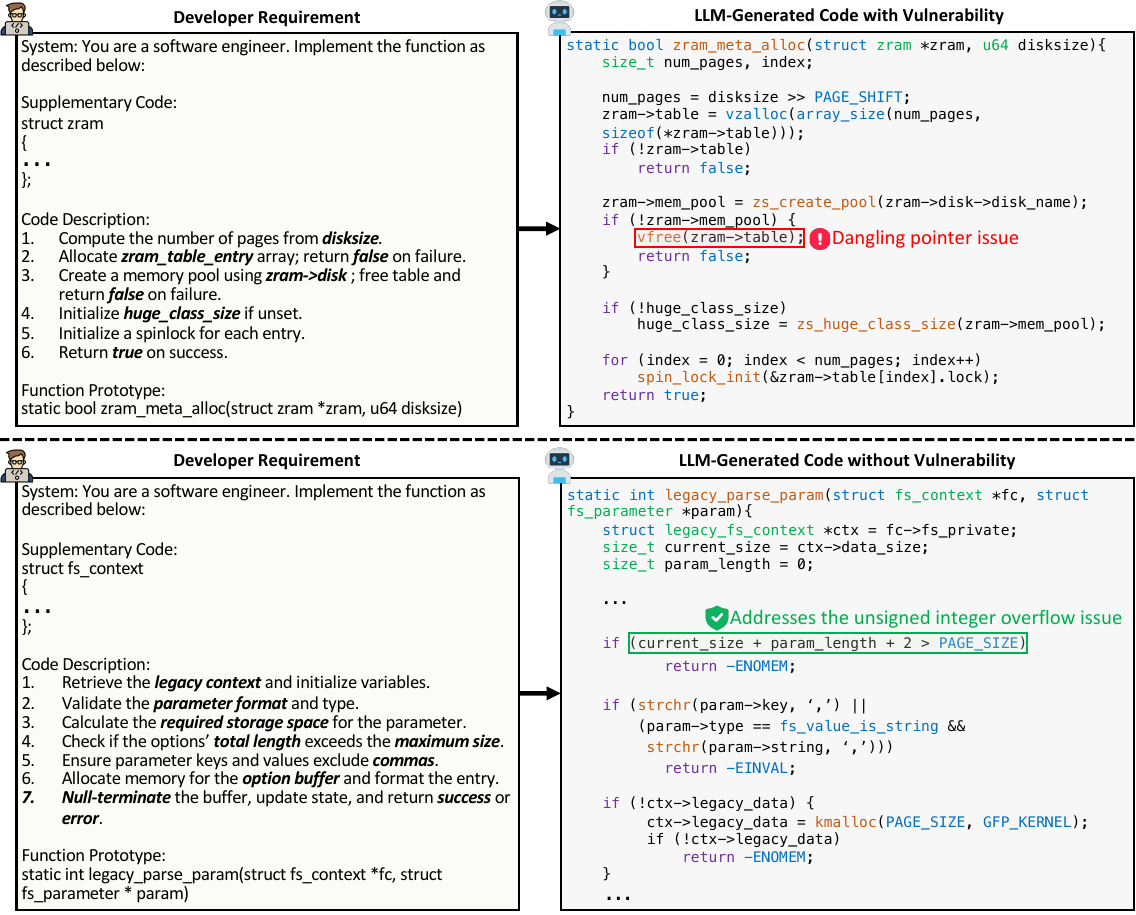}
    \caption{Motivating examples using GPT-4o-2024-11-20. The upper part shows insecure code generation reflecting CVE-2025-21671 (disclosed after the knowledge cutoff). The lower part shows secure code generation for CVE-2022-0185 (disclosed before the cutoff).}
    \label{fig:motiv_3}
\end{figure*}

% \begin{figure*}[t]
%     % \centering
%     % \hspace*{-0.1\linewidth} % adjust this value as needed
%     \includegraphics[width=1.0\textwidth]{figures/Motivation_2_v5.pdf}
    
%     \caption{NEW.}
%     \label{fig:motiv_2}
% \end{figure*}

\noindent\textbf{Knowledge Cutoff.}
LLMs are inherently constrained by a knowledge cutoff, the point in time after which no new data is incorporated into the training corpus due to the substantial cost and overhead of data collection and model training. Consequently, even recent LLM iterations remain bound by fixed cutoff dates. As shown in Table~\ref{tab:codegen_models}, the o3-mini, GPT-4o, and GPT-4o Realtime models have cutoffs in October 2023. Likewise, the built-in code generation models in Copilot’s Codex~\cite{copilot_platform} rely on backbone models (GPT-3.5 Turbo and GPT-4o mini) with cutoffs in September 2021 and October 2023. This implies that LLMs cannot account for updates to target software or newly disclosed CVEs after their cutoff during LLM-assisted development. In contrast, open-source platforms such as Chrome and Linux evolve rapidly; in 2024 alone, the Linux Git repository recorded 75,314 commits, with over 3.6 million lines added and 1.5 million removed~\cite{linux_stats}, alongside 8,093 new Linux-kernel-related CVEs~\cite{Linux2024}. Thus, there is a gap between the static nature of pre-trained LLMs and the continual evolution of software and security issues.

\begin{figure*}[t]
    % \hspace*{-0.05\linewidth} % adjust this value as needed
    \centering
    \includegraphics[width=0.85\linewidth]{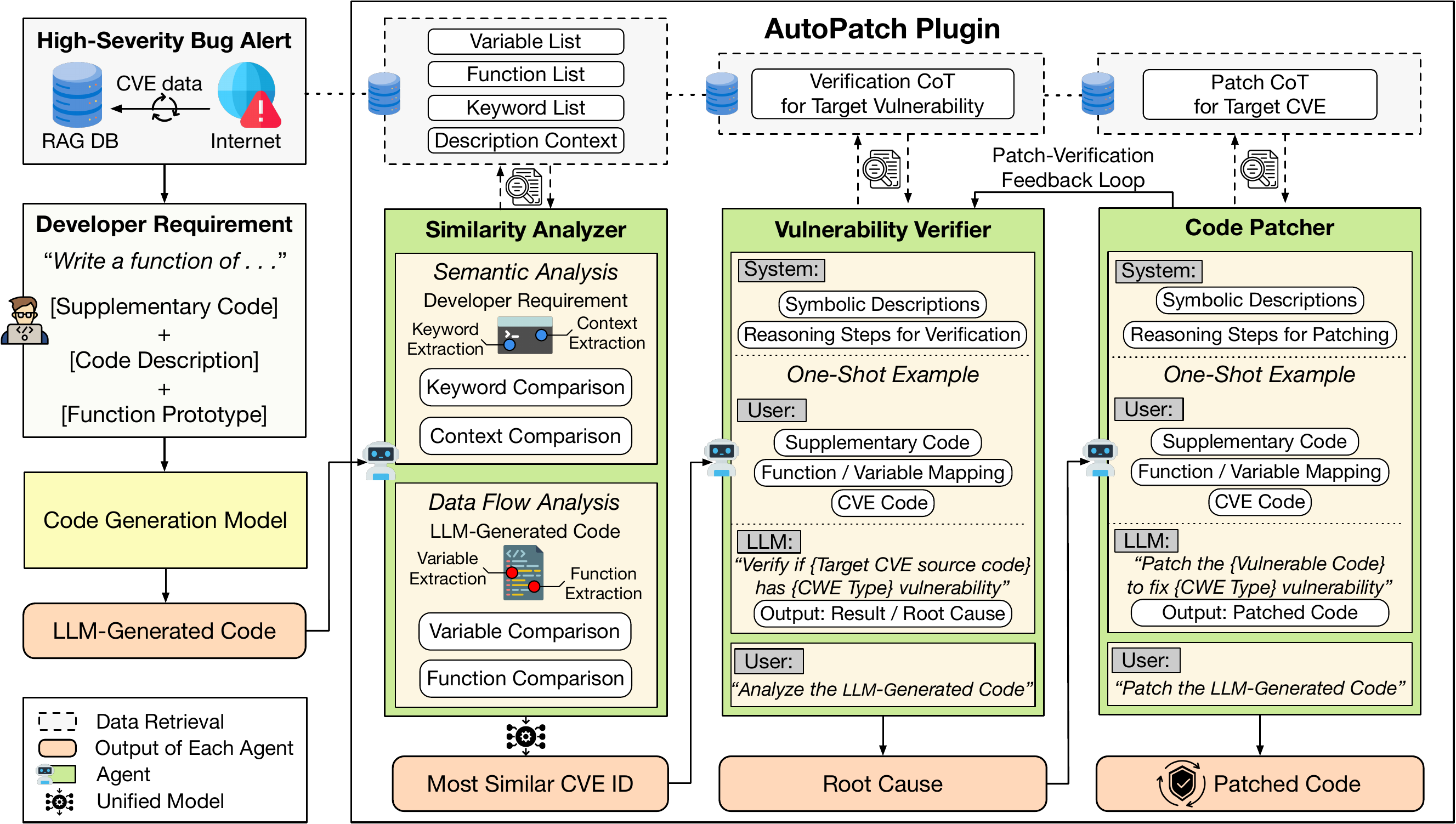}
    
    \caption{The overall architecture of \ourtool{}.}
    \label{fig:system_overview}
    
\end{figure*}
% \subsection{Vulnerable Code Generation}
\noindent\textbf{Vulnerable Code Generation}. 
The static knowledge of LLMs prevents them from reflecting newly disclosed vulnerabilities that are relevant to the developer’s current codebase. When a vulnerability is introduced in libraries, subsystems, or APIs that the ongoing project depends on, any code generated by an LLM cannot account for these updates. As a result, when developers prompt LLMs with context from a recently modified code base, LLMs can produce insecure patterns that have since been identified as vulnerable. For example, as shown in the upper portion of Fig.~\ref{fig:motiv_3}, developers may request a function to initialize metadata and memory structures for a zram device~\cite{zram}, a compressed RAM-based block storage system, by providing a prompt with code snippets referencing existing code bases. However, the code generated by the gpt-4o-2024-11-20 model contains a vulnerability, specifically a dangling pointer issue due to failure to set the pointer to null, which can potentially lead to a use-after-free, CVE-2025-21671~\cite{CVE_2025_21671}, recently disclosed in 2025. Since the model cannot be aware of this newly disclosed vulnerability or its associated secure coding patterns during training, it is unable to avoid the vulnerability, instead mirroring insecure patterns.
% it generates outdated or insecure code that may introduce significant risks. 

\noindent\textbf{Secure Code Generation}. To further examine the security limitations of LLMs in code generation, we conduct an additional experiment using the same gpt-4o-2024-11-20 model. As shown in the lower portion of Fig.~\ref{fig:motiv_3}, developers may request a function that processes individual mount parameters for legacy file systems by converting them into a comma-separated string format. This function is expected to ensure that parameters are of supported types, check for invalid characters such as commas, and prevent buffer overflows. If valid, it appends the key (and the value if present) to a dynamically allocated buffer used during the mount process. This scenario mirrors the vulnerability described in CVE-2022-0185~\cite{CVE_2022_0185}, where the original kernel code miscalculates the buffer length before validating input content, allowing a crafted parameter to overflow the allocated heap buffer. In contrast to the previous zram example involving the newly disclosed CVE, the LLM-generated code in this experiment correctly implements the patched version of the code. This result stems from the fact that CVE-2022-0185 was disclosed before the model's cutoff date, meaning the model had likely been exposed to the patched code artifacts and corresponding discussions of the CVE (or other similar ones) during training. Such artifacts have influenced the LLM to reproduce secure patterns and avoid previously exploited vulnerabilities.

These observations emphasize the need for a systematic framework capable of detecting vulnerabilities in LLM-generated code that are disclosed after a model’s cutoff and generating corresponding secure patches, thereby bridging the gap between the static nature of pre-trained LLMs and the continual evolution of security threats.

% importance of developing a systematic framework to detect and mitigate real-world vulnerabilities in LLM-generated code. By integrating recently disclosed vulnerabilities and corresponding patch information, such a framework can help ensure that generated code remains aligned with the current security state of real-world software, bridging the gap between the static knowledge of code generation models and the dynamic, rapidly evolving nature of open-source software.

% This approach would effectively bridge the gap between the static knowledge of pre-trained models and the dynamic, rapidly evolving nature of open-source software, thereby enhancing both the reliability and safety of AI-assisted software development.

% These observations emphasize the necessity of a systematic approach to regularly identify and mitigate high-severity bugs in real-world code. Such a system would not only improve the security of the LLM-generated code but also bridge the gap between the evolving nature of open-source projects and the static knowledge embedded within LLMs.

% \begin{figure*}[t]
%     % \centering
%     % \hspace*{-0.1\linewidth} % adjust this value as needed
%     \includegraphics[width=1.0\textwidth]{figures/Motivation_2_v5.pdf}
    
%     \caption{GPT-4o-2024-11-20 generated secure code for CVE-2022-0185, correctly addressing the unsigned integer overflow that caused the vulnerability.}
%     \label{fig:motiv_2}
% \end{figure*}

\section{\ourtool{} Design}
\label{sec:design}
\ourtool{} is designed to proactively identify and remediate vulnerabilities, particularly those disclosed after the knowledge cutoff date of code generation models, within the AI-assisted software development workflow, specifically at the stage where developers request code from LLMs.
As shown in Fig.~\ref{fig:system_overview}, \ourtool{} is built on a multi-agent framework comprising three LLM agents tailored for vulnerability detection and patching: the Similarity Analyzer, Vulnerability Verifier, and Code Patcher.
Here,  we first describe the deployment scenario, followed by an introduction to each of these agents.

% \begin{figure*}[t]
%     % \centering
%     % \hspace*{-0.1\linewidth} % adjust this value as needed
%     \includegraphics[width=1.0\textwidth]{figures/variables_functions_prompt_v2.pdf}
    
%     \caption{In this example, the target code refers to the LLM-generated code of CVE-2024-21671 as depicted in Fig.~\ref{fig:motiv}.}
%     \label{fig:variables_functions_prompt}
    
% \end{figure*}

\subsection{\ourtool{} Deployment Scenario}
We consider a typical AI-assisted software development workflow in which developers rely on LLM-integrated IDEs (e.g., Copilot~\cite{copilot_platform} or Cursor~\cite{CursorAI}) to generate code via inline comments or chat-based interactions, with \ourtool{} deployed as a security plugin within the IDE.
Given the LLM's knowledge cutoff, it may produce vulnerable code lacking awareness of recently disclosed CVEs. Our goal is to identify and patch vulnerabilities in LLM-generated code that exhibit similar patterns to previously disclosed vulnerabilities.
When the LLM produces a code snippet, \ourtool{} intercepts it and determines whether it exhibits vulnerabilities similar to recently disclosed CVEs through the Similarity Analyzer and Vulnerability Verifier. If no vulnerabilities are detected, the snippet is forwarded to the IDE and integrated into the existing code base as usual. If vulnerabilities are identified, the Code Patcher generates a corresponding patch and verifies that the revised code preserves the original functionality. The patched snippet is then returned to the IDE for integration.

\subsection{Similarity Analyzer}
% \vspace{-0.15in}

\label{sec:similarity_analzer}
The Similarity Analyzer agent has two key abilities: (i) semantic analysis and (ii) data flow analysis. These abilities work in combination to address two key challenges: detecting code that exhibits \textbf{similar} structures to known vulnerabilities, and identifying \textbf{different} code structures that nonetheless share similar vulnerability patterns. Semantic analysis compares keywords and description contexts from LLM-generated code against known CVEs in our RAG DB, while data flow analysis abstracts variables and functions into symbolic representations for pattern-based matching. To unify these different types of similarity features, we propose a Unified Similarity Model that learns optimal weights over multiple similarity metrics, including keyword, context, variable, and function-level comparisons, to rank the most relevant CVE.

\subsubsection{Semantic Analysis}\mbox{}\\
With semantic analysis ability, the agent calculates a semantic similarity score using two principal strategies: keyword comparison and context comparison.

\noindent\textbf{Keyword Comparison.} In this strategy, keywords are extracted from the developer-provided code description using the top 10,000 most frequently used tags from Stack Overflow~\cite{stackexchange_data}, and compared against keywords stored in the RAG DB, which are derived from CVE code descriptions using the same tag set. To calculate similarity between two keyword sets, the Jaccard similarity score is typically utilized. However, exact keyword matching may miss semantically similar terms with lexical variation. To address this limitation, we incorporate rapidfuzz~\cite{rapidfuzz}, a fuzzy string matching library, and treat two keywords as equivalent if their similarity ratio exceeds 80\%. We then modify the traditional Jaccard formulation by adopting fuzzy set operations, where $\mathbf{\cap_{rf}}$ and $\mathbf{\cup_{rf}}$ represent rapidfuzz-based intersection and union, respectively. This design allows the keyword comparison to remain robust against surface-level variations in naming conventions, while still preserving the discriminative power of keyword overlap. The final similarity score is computed as $J_{kw} = \frac{|A \cap_{rf} B|}{|A \cup_{rf} B|}$.

\noindent\textbf{Context Comparison.} While keyword comparison mainly focuses on matching specific important terms, the context comparison strategy considers the entire semantic context of the given code description. In this strategy, the code description provided by the developer is compared against vulnerable code descriptions stored in our RAG DB. Both descriptions are transformed into high-dimensional vector representations, and cosine similarity is employed to iteratively evaluate their contextual alignment. 

This approach is critical for capturing the functional intent behind the developer's code by aligning it with the descriptions of known vulnerabilities in the RAG DB, even in cases where the exact terminology differs. To achieve this, let $\mathbf{d_d}$ denote the vector representation of the developer's code description and $\mathbf{v_d}$ denote the vector representation of a vulnerable code description retrieved from the RAG DB. 
We employ cosine similarity, which measures the angular distance between their vector representations. The similarity score for context comparison is then computed as $C_{\text{desc}} = \frac{\mathbf{d_d} \cdot \mathbf{v_d}}{\|\mathbf{d_d}\| \|\mathbf{v_d}\|}$.
% follows:
% Then, the similarity score for context comparison is computed as follows:

% \begin{equation}
% C_{\text{desc}} = \frac{\mathbf{d_d} \cdot \mathbf{v_d}}{\|\mathbf{d_d}\| \|\mathbf{v_d}\|}.
% \end{equation}

% \begin{equation} 
% J_{kw} = \frac{|A \cap_{rf} B|}{|A \cup_{rf} B|} 
% \end{equation}

% \noindent\textbf{Context Comparison.} While keyword comparison focuses on matching discrete terms, context comparison captures the broader semantic meaning of the code description. In this strategy, the developer-provided code description is encoded into a high-dimensional vector and compared against vulnerable code descriptions in the RAG DB using cosine similarity. This approach enables alignment based on functional intent, even when exact terminology differs. Let $\mathbf{d}$ and $\mathbf{v}$ denote the vector representations of the developer's and CVE descriptions, respectively. The similarity score is computed as $C_{\text{desc}} = \frac{\mathbf{d} \cdot \mathbf{v}}{\|\mathbf{d}\| \|\mathbf{v}\|}$.
\begin{figure}[t]
    \centering
    \includegraphics[width=1.0\linewidth]{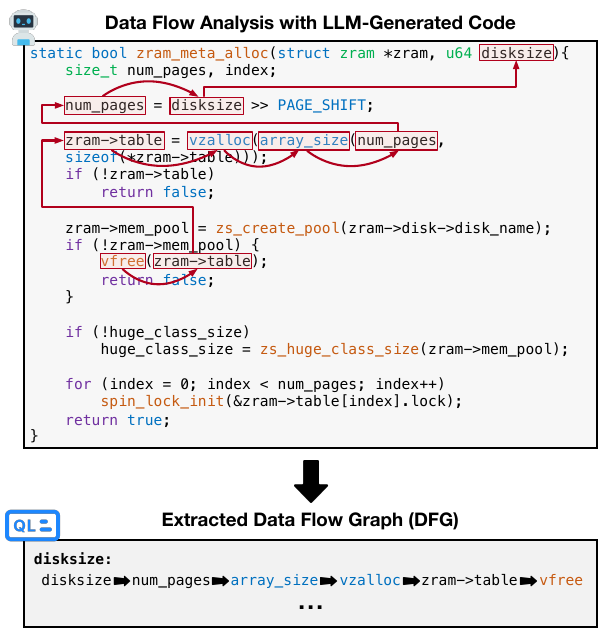}
    \caption{Data Flow Graph (DFG) Extraction.}
    \label{fig:DFG}
\end{figure}

\begin{figure*}[t]
    \centering
    \includegraphics[width=0.85\textwidth]{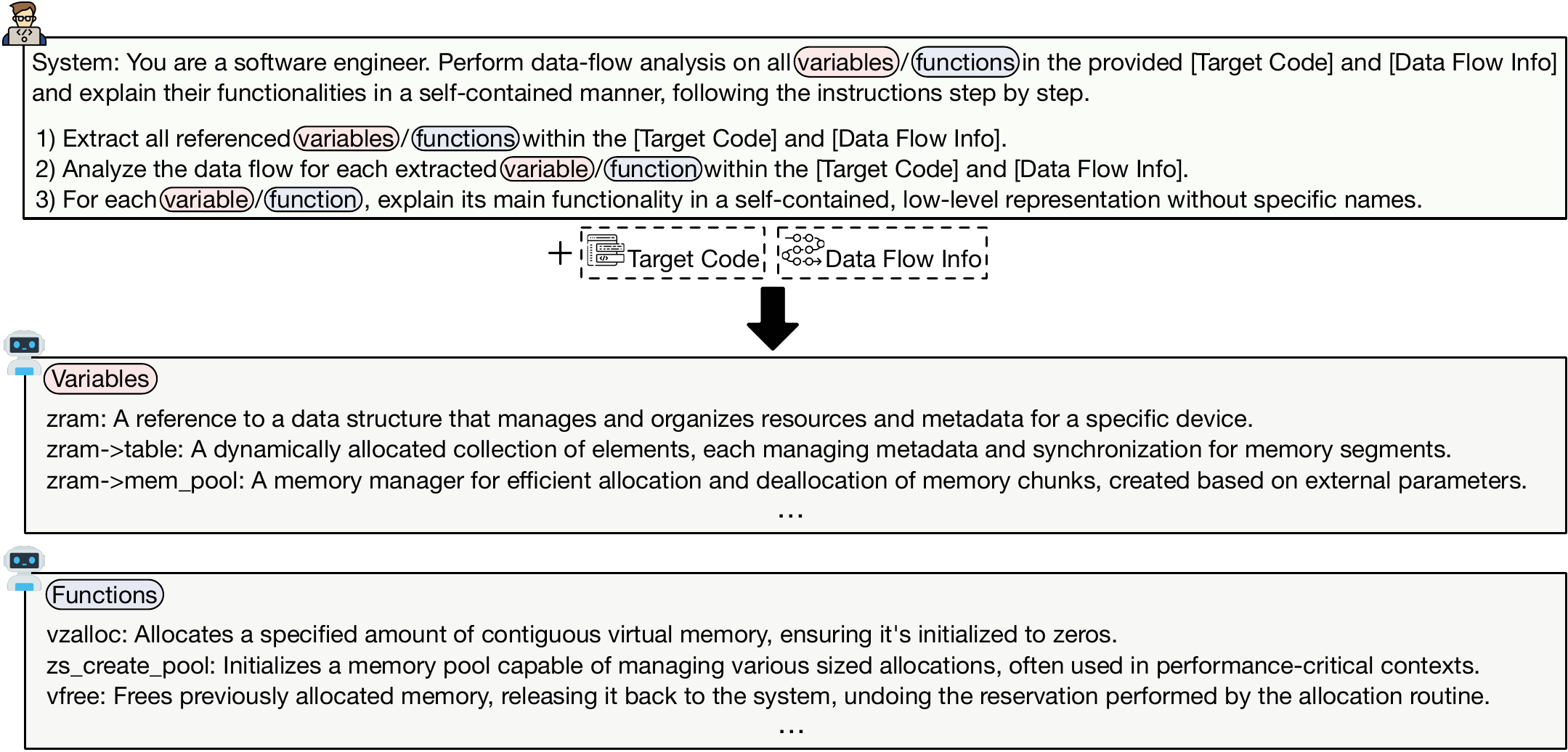}
    \caption{The target code refers to the LLM-generated code of CVE-2024-21671 as depicted in the upper portion of Fig.~\ref{fig:motiv_3}, while the data flow info refers to the structured representation derived from the extracted Data Flow Graph (DFG), shown in Fig.~\ref{fig:DFG}.}
    \label{fig:variables_functions_prompt}
    
\end{figure*}

\subsubsection{Data Flow Analysis}\mbox{}\\

With data flow analysis ability, the agent calculates a data flow similarity score by performing two principal strategies: variable comparison and function comparison. To accurately capture the contextual relationship among variables and functions within the LLM-generated code, we extract a Data Flow Graph (DFG), as shown in Fig.~\ref{fig:DFG}. The DFG provides a structured representation of information propagation across variables and functions, thereby facilitating a deeper semantic understanding of program behavior that extends beyond surface-level syntactic features.

For this purpose, we employ CodeQL~\cite{codeql, codeql_paper}, an open-source semantic code analysis framework developed by GitHub. CodeQL transforms source code into a relational database and represents program elements (e.g., variables, functions, and data-flow dependencies) as queryable entities. By writing declarative queries in CodeQL’s domain-specific language, we can systematically track data flow, identify how values are defined, propagated, or modified, and extract explicit data dependencies between program elements. This capability allows us to construct precise DFGs that capture both direct and indirect variable interactions. 

Subsequently, our agent leverages the extracted DFG to abstract variables and functions in the LLM-generated code into symbolic descriptions by removing literal identifiers. As shown in Fig.~\ref{fig:variables_functions_prompt}, specific variable names are replaced with symbolic roles in the description. This abstraction process emphasizes the inherent roles and relationships of program components rather than their superficial naming conventions.

% \mj{REVISE HERE}As illustrated in Fig.~\ref{fig:variables_functions_prompt}, the agent first extracts variables and functions from the LLM-generated code and abstracts them into symbolic descriptions by removing specific naming details, thereby focusing on their inherent roles rather than literal identifiers.
% \wwc{, similar to the ones shown in Fig.~\ref{fig:verification_and_patch_prompt}: [Vulnerability-Related Variables] and [Vulnerability-Related Functions]}. 
Once these symbolic descriptions are obtained, the Similarity Analyzer compares them with the corresponding representations stored in our RAG DB. To quantify the similarity between the symbolic descriptions of variables and functions, we again employ cosine similarity.

Let $\mathbf{d_v}$ denote the vector corresponding to the symbolic description of variables extracted from the LLM-generated code, and $\mathbf{v_v}$ denote the vector corresponding to the vulnerable variable description retrieved from the RAG DB. The similarity score for variable comparison is defined as $C_{\text{var}} = \frac{\mathbf{d_v} \cdot \mathbf{v_v}}{\|\mathbf{d_v}\| \|\mathbf{v_v}\|}$. 
% follows:
% \begin{equation}
% C_{\text{var}} = \frac{\mathbf{d_v} \cdot \mathbf{v_v}}{\|\mathbf{d_v}\| \|\mathbf{v_v}\|}.
% \end{equation}
% Similarly, let $\mathbf{d_f}$ denote the vector representing the symbolic description of functions extracted from the LLM-generated code, and $\mathbf{v_f}$ denote the vector corresponding to the description of the vulnerable functions retrieved from the RAG DB. The similarity score for function comparison is computed as follows:
Similarly, let $\mathbf{d_f}$ denote the vector representing the symbolic description of functions extracted from the LLM-generated code, and $\mathbf{v_f}$ denote that of the vulnerable functions retrieved from the RAG DB. The similarity score for function comparison is then computed as $C_{\text{func}} = \frac{\mathbf{d_f} \cdot \mathbf{v_f}}{\|\mathbf{d_f}\| \|\mathbf{v_f}\|}$.

% follows:

% \begin{equation}
% C_{\text{func}} = \frac{\mathbf{d_f} \cdot \mathbf{v_f}}{\|\mathbf{d_f}\| \|\mathbf{v_f}\|}.
% \end{equation}

% Let $\mathbf{d}$ denote the vector corresponding to the symbolic description extracted from the LLM-generated code, and $\mathbf{v}$ denote the vector from the RAG DB. The similarity scores for variable comparison and function comparison are computed as $C_{\text{var}} = \frac{\mathbf{d} \cdot \mathbf{v}}{\|\mathbf{d}\| \|\mathbf{v}\|}$ and $C_{\text{func}} = \frac{\mathbf{d} \cdot \mathbf{v}}{\|\mathbf{d}\| \|\mathbf{v}\|}$, respectively. 
In addition to obtaining similarity scores, the most probable mappings from symbolic descriptions to variables and functions are utilized during vulnerability verification and code patching 
(see Sections~\ref{sec:vulnerability_verifier} and~\ref{sec:code_patcher}).

% \begin{equation}
% \tilde{C} = \frac{C + 1}{2} 
% \quad
% \left\{
% \begin{array}{l}
% -1 \mapsto 0, \\
% \phantom{-}0 \mapsto 0.5, \\
% \phantom{-}1 \mapsto 1
% \end{array}
% \right.
% \end{equation}

% \begin{figure*}[t]
%     \centering
%     \includegraphics[width=0.85\linewidth]{figures/combined_reasoning_v3.pdf}
%     \caption{Verification and Patch prompt for LLM-generated code.}
%     \label{fig:verifier_patcher_prompt}
    
% \end{figure*}

\subsubsection{Unified Similarity Model}\mbox{}\\
\noindent\textbf{Unified Similarity Score.} We define a unified similarity score $S$ as a weighted linear combination of the four metrics described above. Let $J_{\text{kw}}$ be the Jaccard similarity on keywords (as defined earlier), and let $\tilde{C}_{\text{t}}$ denote the normalized cosine similarity for $t \in \{\text{desc}, \text{var}, \text{func}\}$, corresponding to the description, variable, and function comparisons, respectively. Each cosine similarity score $C_{\text{t}}$ is normalized using the following transformation:

% $\tilde{C}_{\text{desc}}$, $\tilde{C}_{\text{var}}$, and $\tilde{C}_{\text{func}}$ be the normalized cosine similarities for the descriptions, variables, and functions respectively. Each cosine similarity score, $C_{\text{desc}}$, $C_{\text{var}}$, and $C_{\text{func}}$, is normalized as follows:

% \begin{equation}
% \tilde{C}_{\text{desc,var,func}} = \frac{C_{\text{desc,var,func}} + 1}{2} 
% \quad
% \left\{
% \begin{array}{l}
% -1 \mapsto 0, \\
% \phantom{-}0 \mapsto 0.5, \\
% \phantom{-}1 \mapsto 1
% \end{array}
% \right.
% \end{equation}

\begin{equation}
\tilde{C}_t = \frac{C_t + 1}{2}, \quad t \in \{\text{desc}, \text{var}, \text{func}\}
\quad \text{where } 
\left\{
\begin{array}{l}
-1 \mapsto 0, \\
\phantom{-}0 \mapsto 0.5, \\
\phantom{-}1 \mapsto 1
\end{array}
\right.
\end{equation}

% using $\tilde{C} = \frac{C + 1} {2}$, which maps $-1 \mapsto 0$ and $1 \mapsto 1$ (and $0$ to $0.5$) 

% \begin{figure*}[t]
%     % \centering
%     % \hspace*{-0.18\linewidth} % adjust this value as needed
%     \includegraphics[width=1.0\linewidth]{figures/combined_reasoning_v3.pdf}
    
%     \caption{Verification and Patch prompt for LLM-generated code.}
%     \label{fig:verifier_patcher_prompt}
    
% \end{figure*}

Then, the score $\mathbf{S}$ for a given generated code snippet and a particular candidate CVE code is computed as: \begin{equation}
\label{eq:similarity_score}
S = w_1·J_{\text{kw}} + w_2·\tilde{C}_{\text{desc}} + w_3·\tilde{C}_{\text{var}} + w_4·\tilde{C}_{\text{func}}
\end{equation} where $w_1, w_2, w_3, w_4$ are trainable weights that determine the contribution of each similarity metric. These weights are real-valued parameters that will be learned from training data. A higher unified score $S$ should indicate a greater likelihood that the candidate CVE corresponds to the same vulnerability or issue present in the LLM-generated code.

\noindent\textbf{Pairwise Ranking Loss.} To learn the optimal weights 
$\mathbf{w} = [w_1,\allowbreak w_2,\allowbreak w_3,\allowbreak w_4]$
we employ a pairwise ranking loss on training examples. For each generated code snippet in the training set, we have one known positive CVE (the correct vulnerability that matches the code) and rest of negative CVE candidates (irrelevant vulnerabilities for that code). Let $\textbf{S}^+$ denote the unified similarity score for the positive (correct) CVE and let $\textbf{S}^-$ be the score for a negative (incorrect) candidate. We define the pairwise ranking loss as: \begin{equation} L_{\text{pair}} = \max\Big(0, m - \big(S^+ - S^-\big)\Big) \end{equation} where $\textbf{m}$ is a margin hyperparameter that specifies how much higher the positive score needs to be compared to a negative score for the pair to be considered correctly ranked. This pairwise loss encourages the model to assign a higher unified score to the true CVE than to any incorrect CVEs, with a safety margin. It directly penalizes cases where an irrelevant CVE is ranked too close or higher than the correct one.

\noindent\textbf{Weight Optimization and Final Outcome.}
The weight vector $\mathbf{w}$ is trained to minimize the total pairwise ranking loss across all training examples. We employ gradient-based optimization (i.e., Adam) to adjust the weights in the direction that reduces $L_{\text{pair}}$. The final system takes an LLM-generated code and computes $J_{\text{kw}}$, $C_{\text{desc}}$, $C_{\text{var}}$, and $C_{\text{func}}$ against each CVE candidate in the database and then calculates the unified score $S$ using Equation~\ref{eq:similarity_score}. Then, the CVE with the highest $S$ is returned as the most likely relevant vulnerability.

\subsection{Vulnerability Verifier}
\label{sec:vulnerability_verifier}
% \begin{figure*}[t]
%     % \centering
%     % \hspace*{-0.18\linewidth} % adjust this value as needed
%     \includegraphics[width=1.0\linewidth]{figures/combined_reasoning_v3.pdf}
    
%     \caption{Verification and Patch prompt for LLM-generated code.}
%     \label{fig:verifier_patcher_prompt}
    
% \end{figure*}

\begin{figure*}[t]
    \centering
    \includegraphics[width=0.85\linewidth]{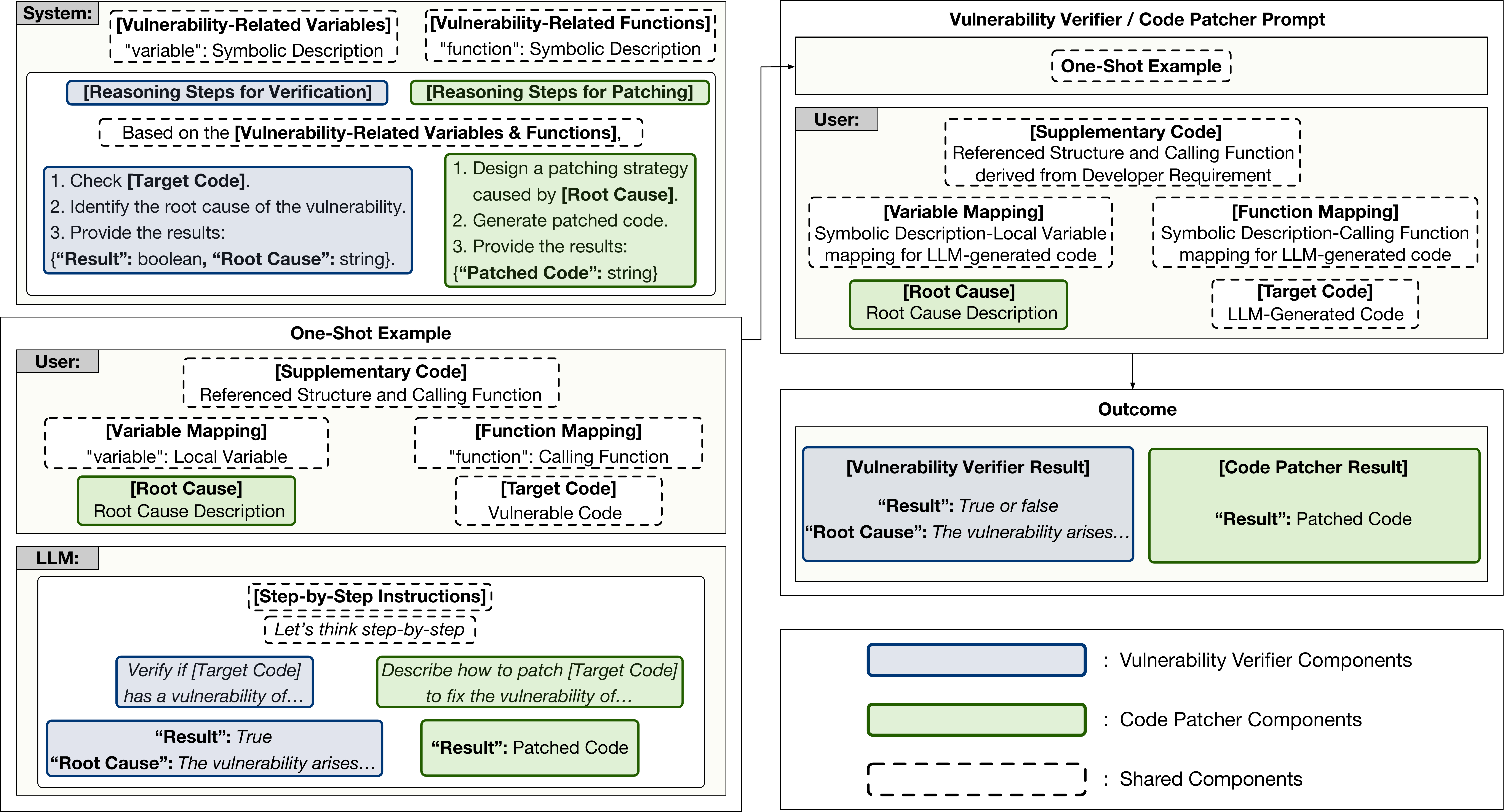}
    \caption{Verification and Patch prompt for LLM-generated code.}
    \label{fig:verifier_patcher_prompt}
    \vspace{-0.1in}
\end{figure*}

Given the most relevant CVE ID along with the mapping from symbolic descriptions to variables and functions provided by the Similarity Agent, the next task is to verify whether the LLM-generated code is vulnerable to a pattern similar to the identified CVE. The Vulnerability Verifier is responsible for this task. This verification step helps avoid unnecessary code patching when no such vulnerability is present. Furthermore, it facilitates the adoption of the patch-verification loop structure, which has been widely employed in Automated Program Repair (APR) studies~\cite{afzal2019sosrepair, xu2020restore}.
For this, The Vulnerability Verifier performs two key tasks: (i) constructing a one-shot example from retrieved CVE metadata and (ii) generating the final verification prompt.

In Fig.~\ref{fig:verifier_patcher_prompt}, the dotted-line box and the blue-colored boxes indicate components related to the Vulnerability Verifier. These prompts follow a typical role-based structure, consisting of three components: system, one-shot example, and user. The system component, shown at the top-left, defines the overall task and provides symbolic descriptions of the variables and functions that play critical roles in triggering the CVE. Notably, the names of variables and functions are abstracted (e.g., "variable\_1" and "function\_1") to enable generalized vulnerability verification. The one-shot example, located at the bottom-left of Fig.~\ref{fig:verifier_patcher_prompt}, serves as an in-context demonstration of correct reasoning, illustrating how each mapped variable and function should be processed to complete the agent's task. It includes CVE data along with mapping information linking the symbolic descriptions to their corresponding code elements in the CVE. Finally, the user component appears on the top-right of Fig.~\ref{fig:verifier_patcher_prompt} and is structured similarly to the user part of the one-shot example, but instead encodes the LLM-generated code that requires vulnerability verification.

% \subsubsection{Vulnerability Verifier}\mbox{}\\
% \label{sec:vulnerability_verifier}
% The Vulnerability Verifier agent constructs a verification prompt to assess whether the LLM-generated code exhibits a vulnerability similar to the CVE identified by the Similarity Analyzer agent. The agent's core abilities are (i) constructing a one-shot example from retrieved CVE metadata and (ii) generating the final verification prompt. In Fig.~\ref{fig:verifier_patcher_prompt}, the dotted-line box and the blue-colored boxes indicate components related to the Vulnerability Verifier agent.

% \input{tables/augmentation_table}
% \input{tables/reimpl_table}

\noindent\textbf{One-Shot Example.}
The one-shot example is dynamically generated from a CVE entry retrieved from the RAG DB. Its user part includes the vulnerable code associated with the CVE, supplementary code (e.g., structure definitions and one-hop calling functions), and the actual mapping from the symbolic descriptions to variables and functions. The LLM response demonstrates how to reason over the symbolic mappings, identify the root cause, and deliver a boolean verdict accompanied by an explanatory rationale.

\noindent\textbf{Verification Prompt.} 
The agent constructs the final verification prompt by concatenating three components. The System component serves as a fixed preamble, instructing the LLM to analyze the provided code for vulnerabilities and identify their root cause. It introduces symbolic descriptions of vulnerability-related variables and functions from the RAG DB and outlines a structured reasoning process for vulnerability verification. The one-shot example, inserted immediately after the System prompt, serves as an in-context demonstration aligned with these symbolic descriptions. Finally, the User component mirrors the structure of the one-shot User input, including relevant structure definitions, one-hop calling functions, symbolic mappings derived from data flow analysis, and the LLM-generated code to be verified. This complete prompt enables the LLM to determine whether a vulnerability exists and explain its root cause. 
We include a real CVE example, CVE-2025-21671, in the Appendix~\ref{sec:appendix_prompt} (see Fig.~\ref{fig:verification_and_patch_prompt}), which corresponds to the same case presented in the motivating example (see the upper portion of Fig.~\ref{fig:motiv_3}).

\subsection{Code Patcher}
\label{sec:code_patcher}
Once a vulnerability and its root cause are identified, the subsequent task involves generating a secure patch and verifying the correctness of the resulting code snippet. The Code Patcher agent is responsible for this task. For this, this agent constructs a structured prompt consisting of three components (system, one-shot example, and user) to guide the LLM in producing a secure patch. While it maintains the same role-based prompt architecture introduced in Section~\ref{sec:vulnerability_verifier}, its objective shifts from verification to patch synthesis. The agent's primary capabilities include: (i) constructing one-shot examples from CVE patch data, (ii) generating patching prompts, and (iii) providing patch feedback to the Vulnerability Verifier agent. In Fig.~\ref{fig:verifier_patcher_prompt}, the dotted-line box and green-highlighted components correspond to modules handled by the Code Patcher agent.

% While maintaining the same role-based structure described in Section~\ref{sec:vulnerability_verifier}, its focus shifts from verification to patch generation. The agent's core abilities are (i) constructing a one-shot example based on CVE patch data, (ii) generating the patching prompt, and (iii) providing patch feedback to the Vulnerability Verifier agent. In Fig.~\ref{fig:verifier_patcher_prompt}, the dotted-line box and the green-colored boxes represent components associated with the Code Patcher agent.

% \input{tables/augmentation_table}
% \input{tables/reimpl_table}

% \begin{table*}[t]
% \centering

% \begin{minipage}[t]{0.48\textwidth}
%     \centering
%     \input{tables/augmentation_table}
% \end{minipage}
% \hfill
% \begin{minipage}[t]{0.48\textwidth}
%     \centering
%     \input{tables/reimpl_table}
% \end{minipage}

% \end{table*}

\noindent\textbf{Patching Prompt.} 
The construction of the one-shot example and patching prompt follows the same structure as in the Vulnerability Verifier (Section~\ref{sec:vulnerability_verifier}) but is adapted to guide patch generation. The user exchange within the one-shot example additionally includes the root cause identified by the Vulnerability Verifier to help the model determine which variables and functions contribute to the vulnerability. The LLM response of the one-shot example demonstrates a reasoning path that leads to a patching strategy and the synthesis of a patched version of the code, rather than a vulnerability verdict. The System prompt is updated to instruct the LLM to generate a secure patch for the given code. As a result, the full prompt enables the generation of a patched variant of the vulnerable LLM-generated code (see Appendix~\ref{sec:appendix_prompt}, Fig.~\ref{fig:verification_and_patch_prompt}).

\noindent\textbf{Patch-Verification Feedback Loop.}
After the Code Patcher agent generates a patched version of the code, vulnerabilities may remain. To ensure the reliability and security of the final output, the system employs a patch-verification feedback loop, executed for a developer-specified number of iterations. In this loop, the code generated by the Code Patcher agent is returned to the Vulnerability Verifier agent, where it performs the same verification process using the previously constructed one-shot example. This cycle continues until either no vulnerability is detected or the maximum number of iterations is reached. Upon completion of the loop, the system outputs the final version of the code, which is considered to be secure by the Vulnerability Verifier agent.

\begin{table}[t]
\footnotesize
\centering
\caption{Trivial and non-trivial code augmentations used in our evaluation.}
\label{tab:code_augmentations}
\vspace{-1mm}
\begin{tabularx}{\linewidth}{c | c | X}
\toprule
\textbf{Type} & \textbf{ID} & \textbf{Description} \\
\midrule

\multirow{5}{*}{Trivial}
& T1 & Add random unreachable codes \\ 
& T2 & Add random codes in comments \\ 
& T3 & Insert whitespaces \\ 
& T4 & Add a useless function \\ 
& T5 & Add next-line characters \\
\midrule

& NT1 & Rename variables and functions to vulnerability-related keywords \\
\multirow{4}{*}{Non-Trivial}
& NT2 & Rename function parameters to vulnerability-related keywords \\
& NT3 & Add a vulnerable library function (e.g., \texttt{strcpy}, \texttt{memcpy}, or \texttt{strcat}) but use it in a safe way \\
& NT4 & Add comments containing keywords related to the vulnerability to the code segment \\
\bottomrule
\end{tabularx}
\vspace{-3mm}
\end{table}

% \footnotesize
% \centering
% \caption{Trivial and non-trivial code augmentations used in our evaluation.}
% \label{tab:code_augmentations}
% \vspace{-1mm}
% \begin{tabularx}{\linewidth}{c | c | X}
% \toprule
% \textbf{Type} & \textbf{ID} & \textbf{Description} \\
% \midrule

% \multirow{5}{*}{Trivial}
% & T1 & Add random unreachable codes \\ 
% & T2 & Add random codes in comments \\ 
% & T3 & Insert whitespaces \\ 
% & T4 & Add a useless function \\ 
% & T5 & Add next-line characters \\
% \midrule

% & NT1 & Rename variables and functions to vulnerability-related keywords \\
% \multirow{4}{*}{Non-Trivial}
% & NT2 & Rename function parameters to vulnerability-related keywords \\
% & NT3 & Add a vulnerable library function (e.g., \texttt{strcpy}, \texttt{memcpy}, or \texttt{strcat}) but use it in a safe way \\
% & NT4 & Add comments containing keywords related to the vulnerability to the code segment \\
% \bottomrule
% \end{tabularx}
% \vspace{-3mm}
% \input{tables/reimpl_table}
\section{Implementation}

\label{sec:Implementation}
We implement a full prototype of \ourtool{}. To rank the most related vulnerabilities, we design a unified model trained with a pairwise loss function using the Adam optimizer. For seamless multi-agent coordination and RAG-enhanced DB retrieval, we utilize LangChain~\cite{langchain}, and adopt PostgreSQL~\cite{postgre} with vector search for entry retrieval, such as variable/function symbolic descriptions, verification/patch reasoning paths, and other details. 

% To reimplement real-world vulnerabilities, we use five popular LLMs: Code Llama, DeepSeek Coder, DeepSeek-R1, GPT-4o, and OpenAI o3-mini. Among these, DeepSeek-R1, GPT-4o, and o3-mini are used for verification and patching due to their reasoning capabilities.

\noindent\textbf{Dataset Collection and Augmentation.} We develop a custom crawler to continuously collect high-severity CVEs from the GitHub Advisory Database~\cite{github_advisory}, Openwall~\cite{openwall}, and the Chromium issue tracker~\cite{chrome_issue_tracker}. From these sources, we collected 75 high-severity CVEs disclosed in late 2024 and 2025, including 57 from the Linux Kernel and 10 from the Chromium project. For each CVE, we extract the developer’s intent from the vulnerable code and convert it into a natural-language prompt. Then, this prompt is used to guide five LLMs, such as Code Llama (13b-instruct), DeepSeek Coder (v2-lite), DeepSeek-R1 (32b), GPT-4o, and OpenAI o3-mini, in generating re-implementation relevant to the CVE. In total, this yields 375 code snippets, enabling us to assess LLMs' ability to reproduce vulnerable patterns and capture structural diversity. Among the models, DeepSeek-R1, GPT-4o, and o3-mini are further used for verification and patch generation due to their reasoning capabilities.

\newcommand{\resultbar}[3]{%
  \begingroup
  \setlength{\fboxsep}{4pt}%
  \setlength{\fboxrule}{4pt}%
  \colorbox{red!40}{\makebox[#1pt][c]{\scriptsize #1}}%
  \colorbox{green!40}{\makebox[#2pt][c]{\scriptsize #2}}%
  \colorbox{orange!40}{\makebox[#3pt][c]{\scriptsize #3}}%
  \endgroup
}

\begin{table}[t]
\centering
\captionsetup{skip=2pt} 
\renewcommand{\arraystretch}{0.8} 
% \footnotesize
\caption{Comparison of Code Reimplementation Accuracy among LLMs.}
\label{tab:reimpl}
\begin{tabular}{@{}l@{\hskip 6pt}c@{\hskip 8pt}c@{}}
\toprule
\textbf{Model} & \textbf{Vuln. Rate} & \textbf{Details} \\
\midrule
Code Llama
& \colorbox{red!15}{\rule[-3pt]{0pt}{11pt} \kern4pt 68.0\% \kern4pt}  
& \resultbar{51}{19}{5} \hspace{2pt} \texttt{/75} \\
DeepSeek Coder
& \colorbox{red!30}{\rule[-3pt]{0pt}{11pt} \kern4pt 80.0\% \kern4pt}   
& \resultbar{60}{10}{5} \hspace{2pt} \texttt{/75} \\
DeepSeek-R1  
& \colorbox{red!40}{\rule[-3pt]{0pt}{11pt} \kern4pt 85.3\% \kern4pt}
& \resultbar{64}{6}{5} \hspace{2pt} \texttt{/75} \\
GPT-4o & \textbf{\colorbox{red!58}{\rule[-3pt]{0pt}{11pt} \kern3.12pt 89.3\% \kern3.12pt}} & 
\resultbar{67}{3}{5} \hspace{2pt} \texttt{/75} \\
o3-mini &\colorbox{red!45}{\rule[-3pt]{0pt}{11pt} \kern4pt \underline{86.7\%} \kern4pt}  
& \resultbar{65}{2}{8} \hspace{2pt} \texttt{/75} \\
\bottomrule
\end{tabular}

% \begin{center}
% \scriptsize
% \colorbox{red!30}{\phantom{X}}~Vulnerable code generation \quad
% \colorbox{green!30}{\phantom{X}}~Safe code generation \quad
% \colorbox{orange!30}{\phantom{X}}~Non-functional code generation
% \end{center}
\begin{center}
\scriptsize
\makebox[\linewidth]{%
  \colorbox{red!40}{\phantom{X}}~Vulnerable \quad 
  \colorbox{green!40}{\phantom{X}}~Secure \quad
  \colorbox{orange!40}{\phantom{X}}~Non-functional%
}
\end{center}
\vspace{-0.25in}
% \vspace{4pt}
% \begin{tabular}{ll}
% \colorbox{red!30}{\phantom{XX}} & Vulnerable code generation \\
% \colorbox{green!30}{\phantom{XX}} & Safe code generation \\
% \colorbox{orange!30}{\phantom{XX}} & Wrong / non-functional code generation \\
% \end{tabular}

\end{table}

To increase variability and rigorously assess the robustness of \ourtool{}, we apply targeted augmentation strategies~\cite{ullah2024llms} tailored to each CVE. For each CVE, we generate additional 75 vulnerable and 75 patched code snippets. The augmentations are divided into trivial transformations, which preserve semantics while perturbing surface representation, and non-trivial transformations, which alter structure in ways that remain functionally correct.

As shown in Table~\ref{tab:code_augmentations}, trivial augmentations include inserting unreachable code fragments, adding random codes in comments, varying whitespace and formatting, defining a useless function, and introducing extraneous newline characters. These manipulations maintain program behavior while introducing syntactic diversity.

Non-trivial augmentations more meaningfully modify code structure while maintaining correctness. Examples include changing variable and function names to vulnerability-related terms (e.g., overflowLen, exploitFlag, xssParser), renaming function parameters to vulnerability-related keywords, and introducing potentially dangerous library functions such as strcpy, strcat, or memcpy in controlled and safe contexts. To further simulate realistic vulnerability-related artifacts, comments seeded with vulnerability-related terminology (e.g., // FIXME(security): potential buffer overflow here or /* WARNING: tainted input sanitized at line 42 */) are injected. These augmentations increase structural diversity without altering the underlying security semantics, thereby enabling more robust evaluation. Our implementations are publicly available at \url{https://github.com/ai-llm-research/autopatch}.

% \input{tables/reimpl_table}
% To account for structural diversity in code, we apply two strategies for each CVE, re-implementation and code augementation. The re-implementation startegy converts the CVE-vulnerable code into developer requirements and instruct LLMs to re-implement based on the requirements. To reimplement real-world vulnerabilities, we use five popular LLMs: Code Llama, DeepSeek Coder, DeepSeek-R1, GPT-4o, and OpenAI o3-mini. Among these, DeepSeek-R1, GPT-4o, and o3-mini are used for verification and patching due to their reasoning capabilities. The augementati

% targeted augmentation strategies for each CVE type, generating an additional 75 vulnerable and 75 patched code snippets. These strategies include renaming local and function parameters using terms typical of other CWE types (e.g., query, freedPtr), aligning parameter names with renamed variables, injecting unreachable code blocks containing potentially vulnerable variable names (e.g., malloc), inserting CWE examples as comments, and adding non-functional whitespace and newline variations.

\section{Evaluation}
\label{sec:evaluation}
We conduct a comprehensive evaluation of \ourtool{} including unified model performance, vulnerability verification, and code patching effectiveness. Also, we analyze the verification and patching performance of \ourtool{} in relation to CWE types and compare its operational cost against traditional fine-tuning approaches.
% \begin{table}[ht]
% \centering
% \small
% \renewcommand{\arraystretch}{1.5}
% \setlength{\tabcolsep}{8pt}

% \begin{tabular}{@{}lcccccc@{}}
% \toprule
% \multicolumn{1}{c}{} 
% & \multicolumn{3}{c}{\textbf{AutoPatch Plugin with Reasoning}} 
% & \multicolumn{2}{c}{\textbf{Existing Techniques}} \\
% \cmidrule(lr){2-4} \cmidrule(lr){5-6}
% & GPT-4o & o3-mini & DeepSeek-R1 & VSP & Baseline \\
% \midrule

% \textbf{CoT Reasoning} \xmarkcircle~F1  
% & 10 & 20 & 30 & 40 & 50 \\

% \textbf{Vulnerability} \cmarkcircle~Acc 
% & 10 & 20 & 30 & 40 & 50 \\

% \midrule

% \textbf{CoT Reasoning} \cmarkcircle~F1  
% & 10 & 20 & 30 & 40 & 50 \\

% \textbf{Vulnerability} \cmarkcircle~Acc 
% & 10 & 20 & 30 & 40 & 50 \\

% \bottomrule
% \end{tabular}
% \caption{Comparison between AutoPatch with CoT Reasoning and existing baselines}
% \end{table}

\begin{table*}[t]

\centering
\begin{minipage}{\textwidth}
\centering
% \small
% \scriptsize
\caption{Comparison of \ourtool{} Plugin Performance During the Verification.}

\caption*{\scriptsize
\makebox[\textwidth][c]{True Positive (TP): Predicted a vulnerability, and a vulnerability existed; CoT was correct.}\\
\makebox[\textwidth][c]{False Positive (FP): Predicted a vulnerability, but no vulnerability existed or CoT was incorrect.}\\
\makebox[\textwidth][c]{False Negative (FN): Predicted no vulnerability, but a vulnerability existed or CoT was incorrect.}\\
\makebox[\textwidth][c]{True Negative (TN): Predicted no vulnerability, and there was no vulnerability; CoT was correct.}
}

\label{tab:verification_and_patch}
\renewcommand{\arraystretch}{1.3}
\setlength{\tabcolsep}{5pt}

\begin{tabular}{@{}llcccccc@{}}
\toprule
\multicolumn{1}{c}{} 
\multirow{2}{*}{\makecell[c]{\hspace*{-2.6cm}\textbf{Task Details}}} & \multirow{2}{*}{\centering\hspace*{0.1cm}\textbf{Metric}}
& \multicolumn{3}{c}{\textbf{AutoPatch with Reasoning Models}} 
& \multicolumn{2}{c}{\textbf{Existing Techniques}} \\
\cmidrule(lr){3-5} \cmidrule(lr){6-7}
\multicolumn{1}{c}{} &  & DeepSeek-R1 & GPT-4o & o3-mini & VSP~\cite{nong2024chain} & Baseline \\
\midrule

\multirow{2}{*}{\makecell[l]{
\textbf{CoT Reasoning} \emarkcircle \\ \hspace*{0.23em} \textbf{Vulnerability} \hspace*{0.15em} \cmarkcircle
}} 
& Accuracy & \cellcolor{green!37}78.54\% & \cellcolor{green!49}\textbf{90.24}\% & \cellcolor{green!40}\underline{80.00\%} & \cellcolor{green!18}46.04\% & \cellcolor{green!20}50.00\% \\
& F1-score & \cellcolor{green!46}\underline{84.17\%} & \cellcolor{green!55}\textbf{92.00}\% & \cellcolor{green!44}83.27\% & \cellcolor{green!15}40.44\% & \cellcolor{green!20}50.73\% \\

\midrule

\multirow{2}{*}{\makecell[l]{
\textbf{CoT Reasoning} \cmarkcircle \\ \hspace*{0.47em}\textbf{Vulnerability} \hspace*{0.26em} \cmarkcircle}} 
& Accuracy & \cellcolor{green!34}75.12\% & \cellcolor{green!47}\textbf{88.29}\% & \cellcolor{green!38}\underline{77.07\%} & \cellcolor{green!9}28.22\% & \cellcolor{green!9}28.22\% \\
& F1-score & \cellcolor{green!42}\underline{81.02\%} & \cellcolor{green!51}\textbf{90.32}\% & \cellcolor{green!40}80.50\% & \cellcolor{green!5}20.77\% & \cellcolor{green!8}27.86\% \\

% \midrule

% \hspace*{1.7em} \textbf{Patched} \hspace*{1.35em} \cmarkcircle
% & Accuracy & \cellcolor{green!35}85.59\% & \cellcolor{green!56}\textbf{95.04}\% & \cellcolor{green!54}\underline{91.30\%} & \cellcolor{green!23}55.56\% & \cellcolor{green!18}46.38\% \\

\bottomrule
\end{tabular}
\end{minipage}
% \caption*{\footnotesize
% True Positive: Predicted a vulnerability, and a vulnerability existed; CoT was correct.\\
% False Positive: Predicted a vulnerability, but no vulnerability existed or CoT was incorrect.\\
% False Negative: Predicted no vulnerability, but a vulnerability existed or CoT was incorrect.\\
% True Negative: Predicted no vulnerability, and there was no vulnerability; CoT was correct.\\
% }
% \caption*{\scriptsize
% \makebox[\textwidth][c]{True Positive (TP): Predicted a vulnerability, and a vulnerability existed; CoT was correct.}\\
% \makebox[\textwidth][c]{False Positive (FP): Predicted a vulnerability, but no vulnerability existed or CoT was incorrect.}\\
% \makebox[\textwidth][c]{False Negative (FN): Predicted no vulnerability, but a vulnerability existed or CoT was incorrect.}\\
% \makebox[\textwidth][c]{True Negative (TN): Predicted no vulnerability, and there was no vulnerability; CoT was correct.}
% }
% \vspace{-0.1in}
\end{table*}

\subsection{Evaluation Environment}
\subsubsection{Code Re-Implementation}
For the code re-implementation experiment, the LLM is configured with a temperature of 0.2 and a top-p value of 0.9. This setup follows common practice in code-generation research: the relatively low temperature promotes stable and consistent outputs, while the moderately high top-p value allows the model to consider a sufficiently diverse set of plausible tokens. Together, these parameters balance determinism with controlled variability, helping the model generate coherent and correct code. The only exception is o3-mini, whose API does not expose temperature or top-p controls; thus, it is used with its default configuration.

\subsubsection{Verification and Patch}
For the verification and patch experiments, \ourtool{} is compared against baseline and VSP~\cite{nong2024chain}. Baseline performs zero-shot prompting and is given only the correct CWE type as context, whereas VSP receives the correct CWE label along with a simple one-shot example for both the verification and patch stages. Both baseline and VSP are implemented using the GPT-4o model, which we adopt because it empirically shows stable performance in both vulnerability verification and patch generation when integrated with \ourtool{}.
% Consistent with the setting adopted by APPATCH~\cite{nong2025appatch}, whose patch strategy also assumes that the explicit vulnerable location is provided to the model rather than discovered autonomously, we treat VSP as operating under an oracle bug-localization regime.
For all methods in this evaluation, the LLM temperature is fixed at 0.0 to ensure fully deterministic behavior. As before, o3-mini is used with its default settings because its API does not allow temperature configuration.

\begin{table}[t]
\centering
\caption{Comparison of \ourtool{} Plugin Performance in the Patching Phase.}
\renewcommand{\arraystretch}{1.25}

\resizebox{\columnwidth}{!}{
\begin{tabular}{ccccccc}
\toprule
\textbf{Metric}
& \multicolumn{3}{c}{\textbf{AutoPatch with Reasoning Models}} 
& \multicolumn{2}{c}{\textbf{Existing Techniques}} \\
\cmidrule(lr){2-4} \cmidrule(lr){5-6}
& DeepSeek-R1 & GPT-4o & o3-mini & VSP~\cite{nong2024chain} & Baseline \\
\midrule

\textbf{Patch acc.}
& \cellcolor{green!43}85.04\% 
& \cellcolor{green!59}\textbf{94.12\%} 
& \cellcolor{green!49}\underline{87.07\%} 
& \cellcolor{green!23}55.56\% 
& \cellcolor{green!18}46.38\% \\

\bottomrule
\end{tabular}
}

\label{tab:patch}
\end{table}

% \begin{table}[h]
% \centering
% % \scriptsize
% \begin{tabular}{lccc}
% \toprule
%  & \textbf{DeepSeek-R1} & \textbf{GPT-4o} & \textbf{o3-mini} \\
% \midrule
% Max loop & 8 & 6 & 10 \\
% Min loop & 1 & 1 & 1 \\
% Avg.\ loop & 1.56 & 1.08 & 1.07 \\
% \bottomrule
% \end{tabular}
% \label{tab:loop}
% \end{table}

\begin{table}[t]
\centering
\caption{Patch-Verification loop statistics for each model.}
\label{tab:loop} 
\begin{tabular}{lccc}
\toprule
 & \textbf{DeepSeek-R1} & \textbf{GPT-4o} & \textbf{o3-mini} \\
\midrule
Max loop & 7 & 6 & 10 \\
Min loop & 1 & 1 & 1 \\
Avg.\ loop & 1.55 & 1.08 & 1.07 \\
\bottomrule
\end{tabular}
\vspace{-0.2in}
\end{table}

\subsection{Unified Model and Code Reimplementation Performance}
Table~\ref{tab:reimpl} shows a comparative analysis of code reimplementation accuracy among various LLMs, based on their vulnerability rates. To assess correctness, we manually verify whether each LLM-generated snippet reproduces real-world CVE vulnerabilities. These annotations serve as the ground truth for training our unified model to identify the most closely matching CVE ID.

The Code Llama model exhibits a 68.0\% vulnerable code generation rate, likely due to higher hallucination and reduced fidelity to the original logic. DeepSeek Coder and DeepSeek-R1 demonstrate higher vulnerability rates of 80.0\% and 85.3\%, respectively, indicating improved structural alignment with ground truth code. Notably, GPT-4o and o3-mini show the highest vulnerability rates, 89.3\% and 86.7\%, respectively, which suggests minimal hallucination and high fidelity in replicating real-world vulnerable patterns.

% Through the reimplementation of CVE code snippets, we initially collected a total of 375 examples. The unified model is designed to perform taint and semantic matching between LLM-generated code and real-world vulnerable code. To prevent overfitting to code snippets that merely look alike, we apply augmentation techniques as described in Section~\ref{sec:Implementation}. The augmentation involves modifying variable names, function call names, and struct declarations to adopt vulnerable-like naming patterns, as well as inserting dummy code structures, thereby preserving semantic integrity while increasing structural diversity. Both vulnerable and patched code snippets are augmented, resulting in an additional 150 code snippets and yielding a final dataset of 525 code snippets.

We train the unified model on the annotated dataset described in Section~\ref{sec:Implementation}, using the Adam optimizer with a pairwise loss function. The data is split into training, validation, and test sets with a ratio of 70:15:15. Training is performed over 500 epochs with a batch size of 12 and a learning rate of 0.005. On the test set, the unified model achieves 91.78\% accuracy in mapping each code snippet to its corresponding CVE ID.

% The primary objective of the unified model is to infer the most appropriate CVE ID corresponding to a given code snippet. The model demonstrates its effectiveness in this task by achieving an accuracy of 90.41\% in identifying real-world vulnerabilities.

% \input{tables/verification_table}
\subsection{\ourtool{} Vulnerability Verifier Performance}

In this section, we evaluate the Vulnerability Verifier agent, which assesses whether LLM-generated code contains a vulnerability and generates a corresponding CoT explanation. Since the collected code snippets exhibit a significant class imbalance, with non-vulnerable examples being relatively sparse, we apply random sampling for each CVE to maintain a 2:1 ratio of vulnerable to non-vulnerable snippets. A prediction is considered correct only if both vulnerability detection and CoT reasoning are accurate. To contextualize the performance of \ourtool{}, we also compare it against two alternative approaches: VSP~\cite{nong2024chain}, which uses a one-shot prompt constructed from a simple CWE-style example relevant to the vulnerability type, and a reasoning-only model, which employs an LLM without any in-context examples. Both utilize the GPT-4o model as the underlying LLM.
% These comparisons demonstrate the effectiveness of our verifier in both accurately identifying vulnerabilities and producing causally grounded explanations.

% The top portion of Table~\ref{tab:verification_and_patch} presents a comparative evaluation of the AutoPatch plugin's performance during the verification phase. \ourtool{} with GPT-4o achieves the highest performance—87.13\% accuracy and 89.52\% F1-score—followed by DeepSeek-R1 and o3-mini, both outperforming traditional methods. The results highlights its robustness in both identifying vulnerabilities and generating accurate reasoning paths. In contrast, VSP achieves only 28.22\% accuracy and 20.77\% F1-score, underscoring its inability to handle the step-by-step reasoning required for real-world vulnerabilities.

% The top portion of Table~\ref{tab:verification_and_patch} presents a comparative evaluation of the AutoPatch plugin's performance during the verification phase. \ourtool{} with GPT-4o achieves the highest performance—87.13\% accuracy and 89.52\% F1-score—followed by DeepSeek-R1 and o3-mini, both outperforming traditional methods. The results highlight its strength in both identifying vulnerabilities and generating accurate reasoning paths. In contrast, VSP achieves only 28.22\% accuracy and 20.77\% F1-score, underscoring its inability to handle the step-by-step reasoning required for real-world vulnerabilities.

Table~\ref{tab:verification_and_patch} presents a comparative evaluation of the \ourtool{} plugin's performance during the verification phase. In the vulnerability-only verification setting, \ourtool{} with GPT-4o achieves the highest performance, 90.24\% accuracy and 92.00\% F1-score, followed by DeepSeek-R1 and o3-mini, both outperforming existing techniques. When jointly evaluating vulnerability detection and CoT reasoning, performance drops across all models due to added complexity. GPT-4o still leads with 88.29\% accuracy and 90.32\% F1-score, demonstrating its robustness in both identifying vulnerabilities and generating accurate reasoning paths. DeepSeek-R1 and o3-mini follow a similar trend with moderate declines. These results emphasize the need for context-aware verification, as traditional methods often fall short in interpreting the semantic complexity of vulnerable code. \ourtool{}, particularly when paired with GPT-4o, demonstrates a robust ability to bridge this gap by enabling accurate and interpretable verification.

Overall, \ourtool{}, particularly when paired with GPT-4o, demonstrates strong capability for accurate and interpretable verification. Existing techniques, such as VSP and the baseline model, perform significantly worse across both vulnerability detection and CoT reasoning tasks. These results underscore the importance of context-aware verification, as conventional methods often struggle to capture the semantic complexity of vulnerable code.

% These results emphasize the importance of incorporating reasoning capabilities into real-world vulnerability verification systems. While existing techniques may identify vulnerabilities, they lack the ability to interpret the semantic depth necessary for analyzing complex, real-world vulnerable codes, a critical requirement for establishing trust and enabling automated patch generation. \ourtool{}, particularly when paired with GPT-4o, demonstrates a robust ability to bridge this gap by enabling accurate and interpretable verification.

\subsection{\ourtool{} Code Patcher Performance}

Among the code snippets identified as vulnerable by the Vulnerability Verifier agent, we employ the Code Patcher agent to generate secure (patched) versions of the code. Table~\ref{tab:patch} presents a comparative analysis of the \ourtool{}'s performance during the patching phase, measuring the accuracy of the generated patches. The results clearly demonstrate that \ourtool{}, when integrated with reasoning models, substantially outperforms existing techniques. GPT-4o achieves the highest patching accuracy at 94.12\%, followed by o3-mini at 87.07\% and DeepSeek-R1 at 85.04\%, showing the strength of advanced language models in capturing and acting upon vulnerability semantics. In contrast, VSP and the baseline model exhibit significantly lower accuracies of 55.56\% and 46.38\%, respectively. This performance gap highlights the limitations of simple CWE-based strategies adopted by existing approaches in handling complex, real-world vulnerabilities.

% Among the code snippets identified as vulnerable by the verifier, we employ the Code Patcher agent to generate secure versions of the code. The bottom portion of Table~\ref{tab:verification_and_patch} provides a comparative evaluation of the \ourtool{} plugin's patching performance, measuring the success rate in remediating identified vulnerabilities. The results clearly demonstrate that \ourtool{} substantially outperforms existing techniques. GPT-4o achieves the highest patching accuracy at 95.04\%, followed by o3-mini at 91.30\% and DeepSeek-R1 at 85.59\%, showcasing the strength of advanced language models in capturing and acting upon vulnerability semantics. In contrast, VSP and the baseline model achieve significantly lower accuracies of 55.56\% and 46.38\%, respectively, underscoring their limitations in handling complex, real-world vulnerability scenarios.

These findings demonstrate the effectiveness of our context-aware patching strategy, which provides models with rich, semantically grounded information about the vulnerable code. Rather than relying on isolated or oversimplified patterns, our approach allows reasoning-capable models to better interpret the structural and functional context of the code, ultimately guiding the generation of more accurate and secure patches.

\subsection{Patch-Verification Feedback Loop}

As shown in Table~\ref{tab:loop}, most vulnerabilities are resolved within very few patch verification iterations, with average loop counts close to one across all models. Although a small number of difficult cases require more iterations, as indicated by the maximum loop values such as 8 for DeepSeek-R1 and 10 for o3-mini, these cases are uncommon. The low minimum and average loop counts show that the feedback mechanism imposes minimal overhead in practice, since the initial patch is typically sufficient. Notably, GPT-4o achieves the lowest average loop count (1.08), suggesting that its first round patches most consistently pass verification without the need for further refinement.

\subsection{Performance Comparison Based on CWE Type}

\begin{table*}[t]
% \centering
% \caption{\centering Comparison of Performance Based on CWE Type.\\
% \centering\footnotesize \textbf{D.S.}: DeepSeek-R1 \quad \textbf{4o}: GPT-4o \quad \textbf{o3-m}: o3-mini \quad \textbf{VSP}: VSP \quad \textbf{Base}: Baseline}
\caption{Comparison of Performance Based on CWE Type.}
\caption*{\footnotesize
\textbf{D.S.}: DeepSeek-R1 \quad
\textbf{4o}: GPT-4o \quad
\textbf{o3-m}: o3-mini \quad
\textbf{VSP}: VSP \quad
\textbf{Base}: Baseline
}

% \caption*{\footnotesize \textbf{D.S.}: DeepSeek-R1 \quad \textbf{4o}: GPT-4o \quad \textbf{o3-m}: o3-mini \quad \textbf{VSP}: VSP \quad \textbf{Base}: Baseline}

\label{tab:cwe_verification}
% \scriptsize
\renewcommand{\arraystretch}{1.2}
\setlength{\tabcolsep}{3pt}

\makebox[\textwidth][c]{%
% TABLE 1
\begin{minipage}[t]{0.325\textwidth} % was 0.325
\centering
\caption*{\hspace*{-0.1cm}\scriptsize \textbf{CoT Reasoning}\emarkcircle~\&~\scriptsize \textbf{Vulnerability}\cmarkcircle}
\resizebox{\linewidth}{!}{%
\begin{tabular}{@{ }llccccc@{ }}
\toprule
CWE & Metric & D.S. & 4o & o3-m & VSP & Base \\
\midrule
\multirow{2}{*}{\textbf{C1}} & Acc & \cellcolor{green!38}\underline{78.6\%} & \cellcolor{green!53}\textbf{92.9}\% & \cellcolor{green!38}\underline{78.6\%} & \cellcolor{green!25}61.2\% & \cellcolor{green!26}62.4\% \\
& F1  & \cellcolor{green!46}\underline{85.0\%} & \cellcolor{green!56}\textbf{94.4}\% & \cellcolor{green!41}83.3\% & \cellcolor{green!26}63.0\% & \cellcolor{green!28}66.0\% \\
\midrule
\multirow{2}{*}{\textbf{C2}} & Acc & \cellcolor{green!42}84.6\% & \cellcolor{green!56}\textbf{94.9}\% & \cellcolor{green!50}\underline{89.7\%} & \cellcolor{green!17}45.0\% & \cellcolor{green!18}47.5\% \\
& F1  & \cellcolor{green!49}89.3\% & \cellcolor{green!60}\textbf{96.0}\% & \cellcolor{green!52}\underline{91.7\%} & \cellcolor{green!10}31.3\% & \cellcolor{green!19}48.8\% \\
\midrule
\multirow{2}{*}{\textbf{C3}} & Acc & \cellcolor{green!29}\underline{72.5\%} & \cellcolor{green!44}\textbf{86.3}\% & \cellcolor{green!29}\underline{72.5\%} & \cellcolor{green!14}38.6\% & \cellcolor{green!18}46.4\% \\
& F1  & \cellcolor{green!40}\underline{79.7\%} & \cellcolor{green!49}\textbf{88.3}\% & \cellcolor{green!34}75.9\% & \cellcolor{green!8}27.8\% & \cellcolor{green!15}41.6\% \\
\midrule
\multirow{2}{*}{\textbf{C4}} & Acc & \cellcolor{green!46}88.9\% & \cellcolor{green!56}\textbf{94.4}\% & \cellcolor{green!50}\underline{91.7\%} & \cellcolor{green!25}60.0\% & \cellcolor{green!18}47.3\% \\
& F1  & \cellcolor{green!48}90.9\% & \cellcolor{green!60}\textbf{95.5}\% & \cellcolor{green!54}\underline{93.3\%} & \cellcolor{green!26}62.1\% & \cellcolor{green!22}54.0\% \\
\bottomrule
\end{tabular}
}
\end{minipage}
\hspace{0.3em}
% TABLE 2
\begin{minipage}[t]{0.325\textwidth}
\centering
\caption*{\hspace*{-0.1cm}\scriptsize \textbf{CoT Reasoning}\cmarkcircle~\&~\scriptsize \textbf{Vulnerability}\cmarkcircle}
\resizebox{\linewidth}{!}{%
\begin{tabular}{@{ }llccccc@{ }}
\toprule
CWE & Metric & D.S. & 4o & o3-m & VSP & Base \\
\midrule
\multirow{2}{*}{\textbf{C1}} & Acc & \cellcolor{green!30}71.4\% & \cellcolor{green!49}\textbf{89.3}\% & \cellcolor{green!36}\underline{78.6\%} & \cellcolor{green!17}45.9\% & \cellcolor{green!18}47.1\% \\
& F1  & \cellcolor{green!37}78.9\% & \cellcolor{green!52}\textbf{91.4}\% & \cellcolor{green!41}\underline{83.3\%} & \cellcolor{green!14}39.5\% & \cellcolor{green!18}47.1\% \\
\midrule
\multirow{2}{*}{\textbf{C2}} & Acc & \cellcolor{green!42}\underline{84.6\%} & \cellcolor{green!57}\textbf{94.9}\% & \cellcolor{green!42}\underline{84.6\%} & \cellcolor{green!5}20.0\% & \cellcolor{green!6}23.8\% \\
& F1  & \cellcolor{green!50}\underline{89.3\%} & \cellcolor{green!61}\textbf{96.0}\% & \cellcolor{green!47}87.0\% & \cellcolor{green!3}13.5\% & \cellcolor{green!6}22.8\% \\
\midrule
\multirow{2}{*}{\textbf{C3}} & Acc & \cellcolor{green!28}69.6\% & \cellcolor{green!43}\textbf{85.3}\% & \cellcolor{green!30}\underline{70.6\%} & \cellcolor{green!7}24.1\% & \cellcolor{green!8}25.5\% \\
& F1  & \cellcolor{green!36}\underline{76.5\%} & \cellcolor{green!48}\textbf{87.4}\% & \cellcolor{green!32}73.7\% & \cellcolor{green!3}13.5\% & \cellcolor{green!5}21.2\% \\
\midrule
\multirow{2}{*}{\textbf{C4}} & Acc & \cellcolor{green!40}83.3\% & \cellcolor{green!49}\textbf{88.9}\% & \cellcolor{green!43}\underline{86.1\%} & \cellcolor{green!14}38.2\% & \cellcolor{green!5}21.8\% \\
& F1  & \cellcolor{green!44}86.4\% & \cellcolor{green!52}\textbf{90.9}\% & \cellcolor{green!48}\underline{88.9\%} & \cellcolor{green!14}39.3\% & \cellcolor{green!10}29.5\% \\
\bottomrule
\end{tabular}
}
\end{minipage}
\hspace{0.3em}
% TABLE 3
\begin{minipage}[t]{0.325\textwidth}
\centering
\caption*{\scriptsize \textbf{Patched} \cmarkcircle}
\resizebox{\linewidth}{!}{%
\begin{tabular}{@{ }llccccc@{ }}
\toprule
CWE & Metric & D.S. & 4o & o3-m & VSP & Base \\
\midrule
\textbf{C1} & Acc & \cellcolor{green!59}\textbf{94.4}\% & \cellcolor{green!58}\underline{94.1\%} & \cellcolor{green!47}88.9\% & \cellcolor{green!29}68.4\% & \cellcolor{green!32}73.7\% \\          
\midrule
\textbf{C2} & Acc & \cellcolor{green!44}85.2\% & \cellcolor{green!53}\textbf{92.3}\% & \cellcolor{green!50}\underline{91.3\%} & \cellcolor{green!7}25.0\% & \cellcolor{green!14}38.5\% \\
\midrule
\textbf{C3} & Acc & \cellcolor{green!39}80.3\% & \cellcolor{green!59}\textbf{94.5}\% & \cellcolor{green!45}\underline{85.2\%} & \cellcolor{green!20}50.0\% & \cellcolor{green!11}33.3\% \\
\midrule
\textbf{C4} & Acc & \cellcolor{green!51}\underline{90.5\%} & \cellcolor{green!61}\textbf{95.2}\% & \cellcolor{green!46}85.7\% & \cellcolor{green!17}45.5\% & \cellcolor{green!15}40.0\% \\
\bottomrule

\end{tabular}

}
\end{minipage}
}
% \caption*{\footnotesize \textbf{C1}: Arithmetic \& Type Errors, \textbf{C2}: Concurrencty Issues, \textbf{C3}: Memory Safety, \textbf{C4}: Validation, Logic, and Resource Handling}
% \caption*{\footnotesize \makebox[\textwidth][c]{
% \textbf{C1}: Arithmetic \& Type Errors \quad
% \textbf{C2}: Concurrency Issues \quad
% \textbf{C3}: Memory Safety \quad
% \textbf{C4}: Validation, Logic, and Resource Handling \\
% }}

\caption*{
\textbf{C1}: Arithmetic \& Type Errors,
\textbf{C2}: Concurrency Issues,
\textbf{C3}: Memory Safety,
\textbf{C4}: Validation, Logic, and Resource Handling}
% \caption*{\footnotesize \makebox[\textwidth][c]{%
% \parbox{1.5\textwidth}{%
% \textbf{C1}: Arithmetic \& Type Errors \quad
% \textbf{C2}: Concurrency Issues \quad
% \textbf{C3}: Memory Safety \quad
% \textbf{C4}: Validation, Logic, and Resource Handling\\}}
% \textbf{D.S.}: DeepSeek-R1 \quad
% \textbf{4o}: GPT-4o \quad
% \textbf{o3-m}: o3-mini \quad
% \textbf{VSP}: VSP \quad
% \textbf{Base}: Baseline \\ 
% }
\vspace{-0.2in}
\end{table*}

\noindent
To further understand the performance of \ourtool{}, we analyze vulnerability-only verification, joint verification with CoT reasoning, and patching results across four representative CWE categories: C1 (Arithmetic \& Type Errors), C2 (Concurrency Issues), C3 (Memory Safety), and C4 (Validation, Logic, and Resource Handling).

Table~\ref{tab:cwe_verification} shows the experimental results of each setting.
In the vulnerability-only verification, \ourtool{} with GPT-4o consistently outperforms all other models, demonstrating strong robustness in detecting diverse types of vulnerabilities. As shown in the first sub-table of Table~\ref{tab:cwe_verification}, it achieves the highest F1-scores in all four categories, including 96.0\% in C2 (Concurrency Issues) and 95.5\% in C4 (Validation, Logic, and Resource Handling), demonstrating its ability to capture both syntactic and semantic vulnerability patterns. DeepSeek-R1 and o3-mini also perform well in C4, achieving F1-scores of 90.9\% and 93.3\%, respectively. These results suggest that our context-aware verification and patching mechanism is particularly effective at surfacing semantic inconsistencies, especially in validation and logic-related vulnerabilities. In contrast, VSP and the baseline model show significantly lower performance across all CWE types. Their F1-scores drop markedly in C2 and C3, with VSP achieving only 31.3\% and 27.8\%, and the baseline model scoring 48.8\% and 41.6\%, respectively. This performance gap suggests that the approaches lacking in-context examples struggle to capture the complex semantic context present in real-world vulnerabilities.

A similar trend is observed in the joint verification setting, which requires both correct vulnerability detection and correct CoT reasoning. \ourtool{} with GPT-4o again leads with the highest F1-scores across all CWE types: 91.4\% in C1, 96.0\% in C2, 87.4\% in C3, and 90.9\% in C4, as shown in the second sub-table of Table~\ref{tab:cwe_verification}. While these scores are slightly lower than in the vulnerability-only setting due to the added reasoning complexity, GPT-4o maintains strong performance, especially in C2, where it effectively handles concurrency issues such as race conditions and improper locking. DeepSeek-R1 and o3-mini also demonstrate reasonable performance in C2 and C4, with o3-mini achieving 88.9\% in C4 and DeepSeek-R1 reaching 89.3\% in C2, reflecting their capacity to reason through thread-sensitive behavior, validation, and logic handling. Conversely, VSP and the baseline model continue to struggle, with F1-scores falling below 40\% across all categories. Their weakest performance is again in C2 and C3, where VSP records only 13.5\% and the baseline scores just 22.8\% and 21.2\%. These results reveal the limitations of approaches that lack context-aware prompting, particularly in complex tasks that require the joint consideration of both CoT reasoning and vulnerability verification.

Lastly, the third sub-table of Table~\ref{tab:cwe_verification} presents patching accuracy across CWE categories. GPT-4o achieves the highest performance, peaking at 95.2\% in C4. DeepSeek-R1 and o3-mini also perform well, maintaining an accuracy of over 80\% across all categories, which reflects their robustness in addressing a wide range of vulnerability patterns. In contrast, VSP and the baseline model show limited effectiveness, particularly in more complex categories such as C2 and C3, where their patching accuracy drops to 25.0\% and 33.3\%, respectively. These results emphasize the critical role of contextual understanding in generating correct vulnerability patches.
\vspace{-0.1in}

\subsection{Cost Comparison: \ourtool{} vs Fine-Tuning}
\begin{figure}[t]
    \centering
    \begin{subfigure}{0.49\linewidth}
        \centering
        \includegraphics[width=\linewidth]{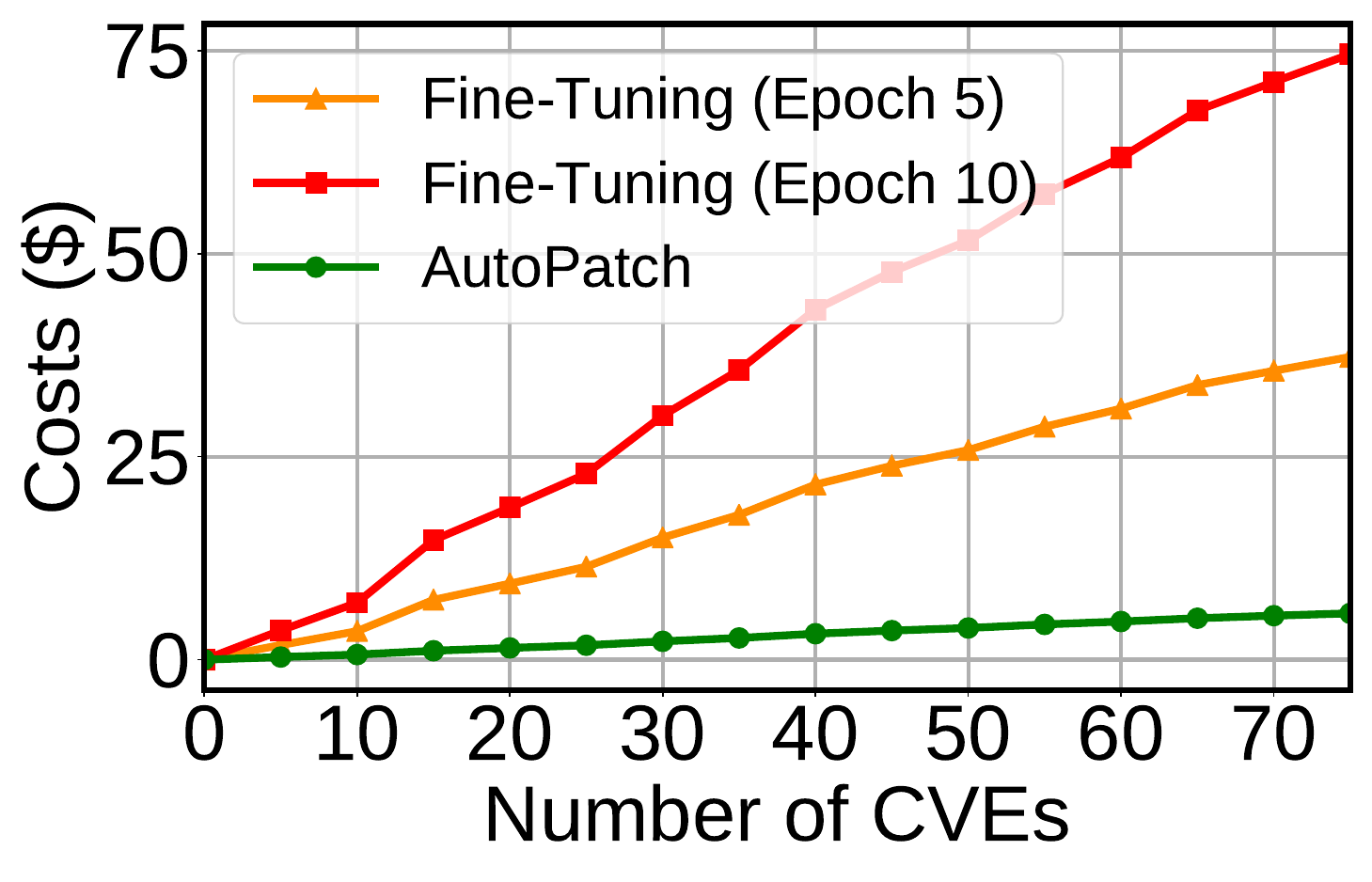}
        \caption{Incremental fine-tuning}
        
        \label{fig:cost_incremental_finetuning}
    \end{subfigure}
    \hfill
    \begin{subfigure}{0.49\linewidth}
        \centering
        \includegraphics[width=\linewidth]{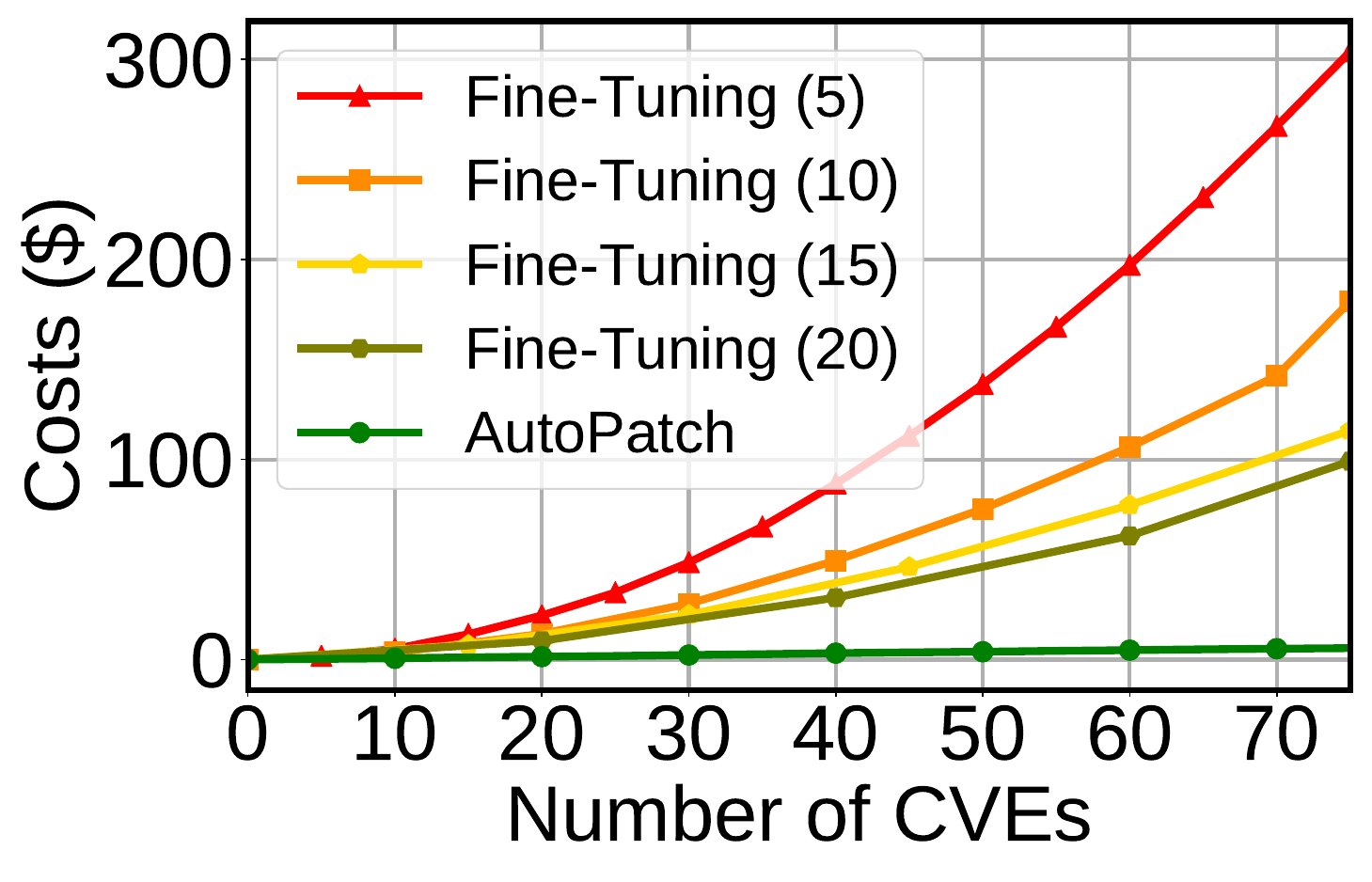}
        \caption{Non-incremental fine-tuning}
        
        \label{fig:cost_non_incremental_finetuning}
    \end{subfigure}
    \caption{Cost comparison for \ourtool{} and fine-tuning.}
    \label{fig:cost_combined}
    \vspace{-0.1in}
\end{figure}

Fig.~\ref{fig:cost_combined} compares the cost of patching CVEs between \ourtool{} and traditional fine-tuning strategies. Here, we use GPT-4o as the base model, as it achieved the highest performance in our prior evaluations. The cost of \ourtool{} is measured based on the information retrieved from the callback mechanism provided by LangChain~\cite{langchain} (i.e., \texttt{get\_openai\_callback}), whereas the cost of the fine-tuning approach is estimated based on OpenAI's per-token API pricing stated by the official website~\cite{openai_platform}.

Both Fig.~\ref{fig:cost_incremental_finetuning} and Fig.~\ref{fig:cost_non_incremental_finetuning} illustrate the cost trends under two common fine-tuning settings: incremental fine-tuning and non-incremental fine-tuning. In the incremental fine-tuning setting, the model is updated sequentially as new CVEs are introduced. While this setting avoids retraining from scratch, it still incurs repeated training overhead, resulting in linearly increasing costs, as shown in Fig.~\ref{fig:cost_incremental_finetuning}. Following standard practices in previous studies~\cite{jan2024multitask,chen2025codesteer}, using 5 or 10 epochs results in costs of \$37.3 and \$74.6, respectively, when patching 75 CVEs. In contrast, the non-incremental fine-tuning setting retrains the model from scratch using all CVE data seen so far, leading to quadratically increasing costs. For instance, fine-tuning every 20 CVEs costs \$99.1, while fine-tuning every 5 CVEs drives the cost up to \$303.8 by the time 75 CVEs are processed, as shown in Fig.~\ref{fig:cost_non_incremental_finetuning}.

\ourtool{}, on the other hand, eliminates the need for any model parameter updates. Instead, it performs lightweight RAG database entry updates when new high-severity CVEs are disclosed. Our evaluation shows that this approach results in a nearly constant operational cost, peaking at only \$5.7, regardless of the number of CVEs processed. This minimal and stable cost profile significantly reduces the computational and operational burden typically associated with maintaining secure, up-to-date vulnerability detection systems.
Compared to \ourtool{}, incremental fine-tuning with 10 epochs is approximately 1,209\% more expensive, non-incremental fine-tuning at 20-CVE intervals is 1,639\% more expensive, and non-incremental fine-tuning at 5-CVE intervals is 5,230\% more expensive. These results show \ourtool{}'s exceptional cost-efficiency and scalability, demonstrating its practicality and sustainability as an alternative to fine-tuning-based approaches for real-world vulnerability detection and patching.

% Figure~\ref{fig:cost_incremental_finetuning} shows the cost under incremental fine-tuning, where the model is updated sequentially as new CVEs are introduced. While this avoids retraining from scratch, it still incurs repeated training overhead, resulting in linearly increasing costs. Existing studies commonly adopt 5 or 10 epochs~\cite{jan2024multitask,krupkina2024badgpt,weyssow2023exploring,chen2025codesteer}, and under this setting, the cost reaches \$37.3 (Epoch 5) and \$74.6 (Epoch 10) when handling 75 CVEs.

% Figure~\ref{fig:cost_non_incremental_finetuning} reflects the non-incremental fine-tuning setting, where the model is retrained from scratch using all CVE data seen so far. This leads to quadratically increasing costs. For example, fine-tuning at an interval of 20 CVEs data already costs \$99.1, while using an interval of 5 on 75 CVEs data results in a cost of \$303.8.

% In contrast, the \ourtool{} plugin performs only RAG database entry updates when new high-severity CVEs are disclosed, without any model updates. Its cost remains nearly constant regardless of CVE count, peaking at only \$5.7. Compared to , incremental fine-tuning with 10 epochs is approximately 1,209\% more expensive, non-incremental fine-tuning at an interval of 20 CVEs is 1,639\% more expensive, and non-incremental fine-tuning at an interval of 5 CVEs is 5,230\% more expensive. These results show \ourtool{}'s cost-efficiency and scalability, making it a practical alternative to fine-tuning-based approaches in real-world. 

\section{Discussion and Limitations}
\label{sec:discussion}

% \ourtool{} leverages a retrieval-augmented generation (RAG) framework over a CVE-based knowledge base to automatically verify and patch vulnerable code. While its design allows for generalization beyond the original application context, \ourtool{} is fundamentally limited to known vulnerabilities. Specifically, it relies on prior examples of CVEs and their associated patches to reason about and repair new code snippets. As a result, it cannot directly detect or repair vulnerabilities that have no precedent in the knowledge base, such as zero-day vulnerabilities or novel exploit patterns. While leveraging LLM assistance to discover unknown vulnerabilities is an interesting research direction, it falls outside the scope of this work. The primary contribution of \ourtool{} lies in automating the knowledge integration pipeline, thereby reducing the window of exposure for high-severity vulnerabilities once disclosed. 

\ourtool{} leverages a retrieval-augmented generation (RAG) framework over a CVE-based knowledge base to automatically verify and patch vulnerable code. While its design allows for generalization beyond the original application context, as shown in Appendix~\ref{sec:autopatch_demonstration}, \ourtool{} is fundamentally limited to known vulnerabilities. Specifically, it relies on prior examples of CVEs and their associated patches to reason about and repair new code snippets, and thus cannot directly detect or repair vulnerabilities with no precedent in the knowledge base, such as zero-day vulnerabilities or novel exploit patterns. Nevertheless, our tool is not only applicable to LLM-generated code but is also readily extendable to existing codebases, as its vulnerability detection and patching pipeline does not assume any dependence on code origin. Moreover, \ourtool{} can still generalize to identify related vulnerability patterns even when different code implementations are provided. While the current dataset is relatively limited and partially manually annotated, we anticipate that by continuously expanding our knowledge base with new vulnerabilities, \ourtool{} can incrementally broaden its detection coverage and improve robustness, while maintaining transparency and compatibility as a plugin for emerging AI-driven development tools. The primary contribution of \ourtool{} lies in automating this knowledge integration pipeline, thereby reducing the window of exposure for high-severity vulnerabilities once disclosed.

%\ww{While AutoPatch is designed with developer-centric usability in mind-intentionally avoiding heavyweight static analysis tools such as AST parsing or taint tracking—future work could explore extending its capabilities toward more security-focused applications. This includes integrating traditional static analysis techniques to enhance vulnerability localization, taint flow validation, and improve the retrieval of related vulnerabilities from the knowledge base by enabling more precise and semantically rich matching. By combining the flexibility of LLM-guided patching with the rigor of static program analysis, AutoPatch could bridge the gap between developer-assistive tools and security-grade automated repair systems, enabling more comprehensive detection and mitigation of complex, data-dependent vulnerabilities.}

An additional limitation in the current \ourtool{} is the relatively small number of Chrome-related CVEs. This limitation primarily arises from Chrome's vulnerability disclosure policy: high-severity vulnerabilities are not released to the public immediately, as a delay is enforced to ensure sufficient time for patch deployment. Despite this challenge, we mitigate the issue by regularly crawling Chrome's issue tracker, which enables us to identify and incorporate 10 relevant CVEs into our dataset. While the coverage remains incomplete due to disclosure constraints, this approach demonstrates the feasibility of extending \ourtool{} to additional platforms as more vulnerabilities become available.

% \ms{LLM-generated code뿐 아니라, 이미 존재하는 코드에 대해서도 사용할 수 있는지, 그 확장성에 대해서 간략하게 언급}

% While

% is primarily because Chrome vulnerabilities, particularly those classified as high-severity, are not made publicly available immediately. These vulnerabilities often undergo a delayed disclosure process to allow time for patch deployment. Nevertheless, by regularly crawling Chrome’s issue tracker, we were able to identify and include 10 CVEs in our dataset.

\section{Related Work}
\label{sec:related_work}

\subsection{LLM-based Vulnerability Detection}
In the domain of LLM-based vulnerability detection, recent surveys \cite{ding2024vulnerability, khare2025understanding, liu2024vuldetectbench} underscore the strong potential of LLMs to significantly improve automated vulnerability analysis.
Prior works have explored frameworks that integrate LLMs with external context to enhance detection accuracy~\cite{lu2024grace, du2024vul, sheng2024lprotector, li2024cleanvul}. For example, GRACE \cite{lu2024grace} enhances LLM-based detection by incorporating graph-structured code representations for fine-grained vulnerability localization, and Vul-RAG \cite{du2024vul} improves detection accuracy via a knowledge-level RAG framework. However, despite these advances, such approaches primarily target generic vulnerability categories (e.g., CWE types) and rely solely on the LLM to identify flaws. In contrast, \ourtool{} incorporates a promptable CVE-level RAG database and employs a unified similarity model to achieve more precise CVE matching. Moreover, \ourtool{} not only detects vulnerabilities but also provides explanations and automatically generates patches for the identified issues.

\subsection{LLM-based Code Repair}
Alongside our approach, several attempts have explored using LLMs for automated software patching. For example, Nong et al. introduced a vulnerability-semantics-guided CoT approach (VSP)~\cite{nong2024chain}, which improved the detection of vulnerabilities (both a given type and unknown types) and the generation of correct patches, outperforming several baselines. While VSP enhances prompting through semantic guidance, it lacks a deep reasoning process for vulnerability analysis. Instead, it primarily optimizes prompt engineering based on semantic information from a given code snippet. In contrast, \ourtool{} combines semantic analysis and data flow analysis with prior CVE data to guide LLMs toward more context-aware vulnerability analysis.

APPATCH proposed an LLM-based automated patching framework~\cite{nong2024automated}. It applies the prompting techniques from VSP for patching and engages an LLM in adaptive reasoning steps to fix code. However, APPATCH has practical usability constraints, particularly in its reliance on precisely identifying the vulnerable line of code as an input to the model. While this assumption may be feasible for code snippets with known vulnerabilities with predefined locations (e.g., those found through static analysis or CVE reports), it is impractical for detecting and patching unknown or newly emerging security flaws. \ourtool{} is not limited to code snippets with known vulnerable lines, as it allows LLMs to actively utilize patterns learned from previous CVEs. 

%ThinkRepair
ThinkRepair is a framework that leverages LLMs with CoT prompting to generate bug fixes with reasoning~\cite{yin2024thinkrepair}. It operates in two phases: first, it constructs a knowledge base of buggy and fixed code annotated with reasoning steps; then, it uses this pool for few-shot prompting to repair new code. Compared to ThinkRepair, our approach further enhances LLM guidance by incorporating variable and function mappings to strengthen the connection between the generated code and the knowledge base.

\section{Conclusion}
\label{sec:conclusion}
We present \ourtool{} plugin, a multi-agent framework that secures LLM-generated code through retrieval-augmented vulnerability detection and patching. We reimplement 525 code snippets based on 75 high-severity, real-world CVEs using five popular LLMs to evaluate our system. Among them, GPT-4o shows the best performance, achieving an F1-score of 90.3\% in vulnerability verification and 94.1\% in patching, particularly excelling in concurrency-related issues and validation and logic handling issues. Compared to traditional fine-tuning approaches, \ourtool{} is significantly more efficient. Compared to traditional fine-tuning approaches, \ourtool{} demonstrates significantly greater efficiency. Specifically, incremental fine-tuning with 10 epochs incurs approximately a 1,209\% higher cost, while non-incremental fine-tuning at 5-CVE intervals results in a 5,230\% increase. These results show that \ourtool{} provides an effective solution for adapting LLMs to newly disclosed vulnerabilities.

\ifCLASSOPTIONcaptionsoff
  \newpage
\fi



%\bibliographystyle{IEEEtran}

% \end{thebibliography}
\twocolumn 
\bibliographystyle{IEEEtran}
\bibliography{bib}

\newpage
\appendix

\subsection{\ourtool{} Demonstration}

\label{sec:autopatch_demonstration}

\begin{figure*}[t]
    \centering
    % \hspace*{-0.05\linewidth}
    \includegraphics[width=0.8\linewidth]{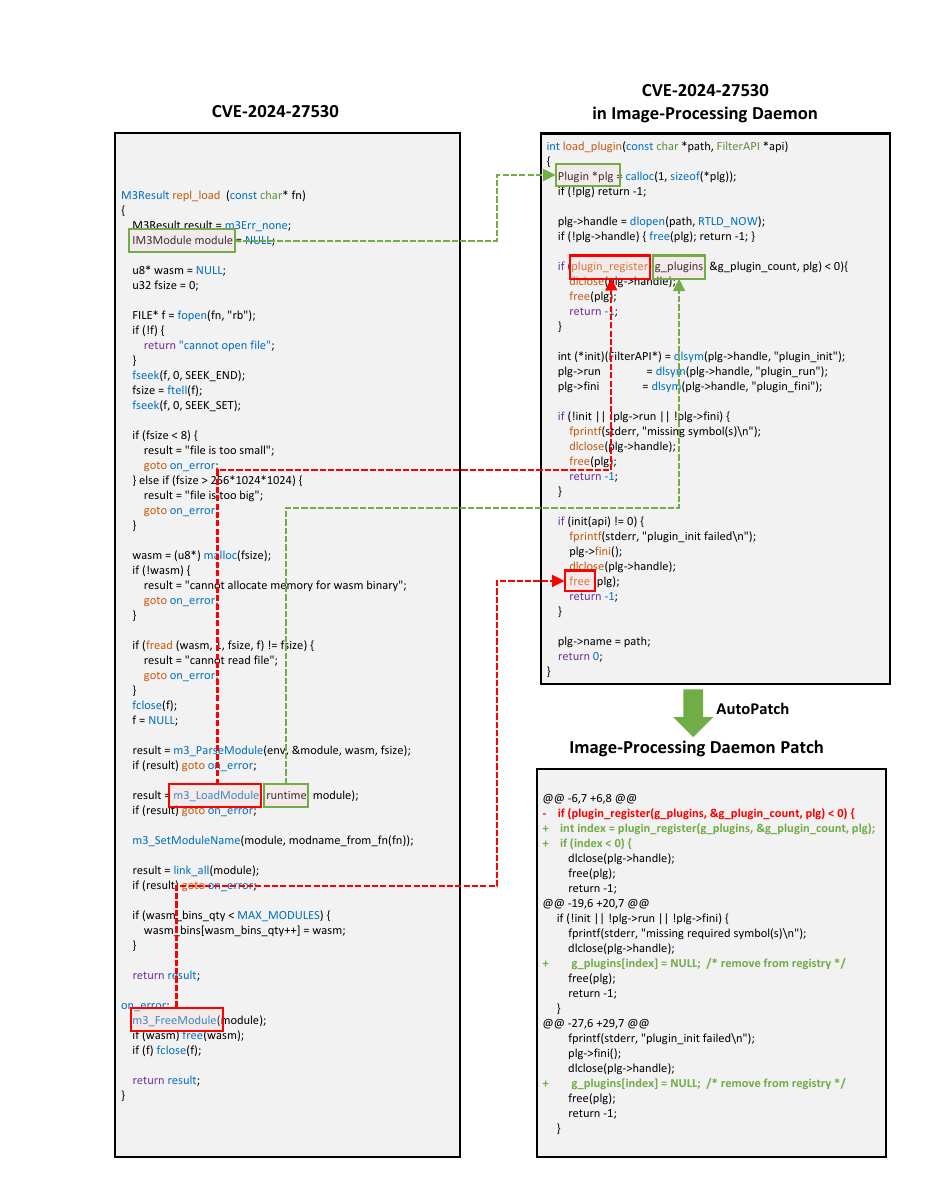}
    \caption{\ourtool{} with \texttt{load\_plugin} function of Image-Processing Daemon.}
    \label{fig:image_daemon}
\end{figure*}

To demonstrate how \ourtool{} verifies and patches vulnerable code, we implemented a simple Image-Processing Daemon that accepts RGB/RGBA image buffers from local clients, processes them through a configurable pipeline of dynamically loaded filter plug-ins (shared objects), and returns the transformed image. Fig.~\ref{fig:image_daemon} illustrates the moment \ourtool{} intervenes as a developer leverages an LLM to implement the \texttt{load\_plugin} function---responsible for loading plug-in files. The LLM-generated \texttt{load\_plugin} is vulnerable to a Use-After-Free, closely resembling CVE-2024-27530, a vulnerability in the WebAssembly interpreter (wasm3) where a \texttt{module} is freed without being properly unregistered from the global module list managed within \texttt{runtime}.

Through semantic analysis and data flow analysis, \ourtool{} queries its RAG-backed database and identifies \texttt{load\_plugin} as being semantically similar to the vulnerable function in CVE-2024-27530. Data flow analysis further maps key variables and functions to aid verification. In this case, the mappings are:  
\begin{itemize}

  \item \textbf{Variables:} \texttt{plg} $\rightarrow$ \texttt{module}, \texttt{g\_plugins} $\rightarrow$ \texttt{runtime}
  \item \textbf{Functions:} \texttt{plugin\_register} $\rightarrow$ \texttt{m3\_LoadModule}, \texttt{free} $\rightarrow$ \texttt{m3\_FreeModule}
\end{itemize}

Along with these mappings, \ourtool{} retrieves both the verification CoT and patch CoT from the database entry for CVE-2024-27530. It then proceeds to verify and patch the Use-After-Free vulnerability in \texttt{load\_plugin} by ensuring that the global list (\texttt{g\_plugins}) is cleared when the plugin (\texttt{plg}) is freed.

\newpage

\newpage

\onecolumn 
\section{Prompt Example}
\subsection{\ourtool{} Prompts}
\label{app:prompt}
\lstdefinelanguage{Prompt}{
    morekeywords={},
    sensitive=True,
    morecomment=[l]{//},
}

\lstset{
  basicstyle=\ttfamily\footnotesize,
  breaklines=true,
  columns=fullflexible,
  keepspaces=true,
  frame=single,
  language=Prompt,
  keywordstyle=\bfseries\color{blue!60!black},
  commentstyle=\itshape\color{gray},
  captionpos=b,
  xleftmargin=2mm,
  xrightmargin=2mm,
}

\begin{lstlisting}[caption={Prompt used for semantic analysis. This prompt analyzes the main functionality of \texttt{[Target Code]}, the LLM-generated code, for later comparison.}]
**Role**: You are an expert software engineer without any software security knowledge. Your goal is to analyze [Target Code] and provide a self-contained summary of its functionality. 
---
**Task Overview**:
Perform the followings step by step and show the reasoning in each step. You are not aware of software security information so DO NOT deduce any security implication on any step. Start answering with "Let's think step-by-step."
You must:
1. Analyze the main functionality of [Target Code].
2. Explain the main functionality of [Target Code] in a self-contained low-level representation. The explanation must be general that it must not include any variable or function names.
3. Finally, provide the self-contained explanation of [Target Code] in the following schema, including the leading and trailing "```json" and "```" 
```json
{
        "result": string // the self-contained explanation of [Target Code] (Step 2)
}
```
\end{lstlisting}

\begin{lstlisting}[caption={Prompt used for function description generation. This prompt first extracts the functions referenced within \texttt{[Target Code]}, the LLM-generated code. The extracted functions are then analyzed to generate descriptions of their functionalities within \texttt{[Target Code]}.}]
**Role**: You are an expert software engineer. Your goal is to perform data-flow analysis on each of the functions referenced by the user-provided [Target Code] and provide entirely self-contained explanations of the functions' functionalities in [Target Code]. 
---
**Task Overview**:
Perform the followings step by step and show the reasoning in each step. Start answering with "Let's think step-by-step."
1. Extract all the referenced functions (including function-like macro with parentheses) within [Target Code].
2. For each extracted function, trace its **data flow** within [Target Code].  
- Use the [Data Flow] (format: "source variable/function" => destination variable/function list) to track how the function is used, how its output is propagated, and how it interacts with other variables or functions.
- If the function is in [Supplementary Code], you can use it to understand the data flow.
3. For each function, generate a **low-level, self-contained explanation** of its functionality.  
- The explanation MUST include:  
    - The role of the function's inputs (where they originate and how they are validated or transformed).  
    - The internal operations (e.g., logical branching, arithmetic, memory management, data structure manipulation, iteration, synchronization).  
    - How the function interacts with external state or other components (e.g., modifies buffers, updates counters, signals errors).  
    - The function's final outcome (e.g., initializes a resource, validates conditions, propagates data, releases memory, introduces risks).  
- The explanation MUST NOT reference specific variable/function names. Instead, describe their roles in **abstract technical terms** (e.g., "a memory buffer holding intermediate graphical state", "a counter that governs iteration termination").
- Each explanation MUST be **self-contained** so it can be understood in isolation, without looking at [Target Code].  
4. Finally, provide the self-contained function functionality explanations in valid JSON format with the following schema, including the leading and trailing "```json" and "```" 
```json
{
        "<function_1_name>": string,  // the self-contained explanation of function_1
        "<function_2_name>": string,  // the self-contained explanation of function_2
        ...
        "<function_n_name>": string  // the self-contained explanation of function_n
}
```
\end{lstlisting}

\begin{lstlisting}[caption={Prompt used for variable description generation. This prompt first extracts the variables referenced within \texttt{[Target Code]}, the LLM-generated code. The extracted variables are then analyzed to generate descriptions of their functionalities within \texttt{[Target Code]}.}]
**Role**: You are an expert software engineer. Your goal is to perform data-flow analysis on each of the variables referenced by the user-provided [Target Code] and provide entirely self-contained explanations of the variables' functionalities in [Target Code]. 
---
**Task Overview**:
Perform the followings step by step and show the reasoning in each step. Start answering with "Let's think step-by-step."
1. Extract all the referenced variables within [Target Code].
2. For each extracted variable, trace its **data flow** within [Target Code].  
- Use the [Data Flow] (format: "source variable/function" => destination variable/function list) to track how the variable is initialized, transformed, passed to functions, or conditionally manipulated.  
3. For each variable, generate a **low-level, self-contained explanation** of its functionality.  
- The explanation MUST include:  
    - The origin of the variable (input, derived from another variable, returned from a function, etc.).  
    - The operations performed on it (arithmetic, logical checks, memory management, iteration, dereferencing, etc.).  
    - How it interacts with other variables or functions (dependencies, propagation, transformations).  
    - The final role or outcome (what state it contributes to, what it enables, what risk it introduces).  
- The explanation MUST NOT reference specific variable/function names. Instead, describe their roles in **abstract technical terms** (e.g., "a memory buffer holding intermediate graphical state", "a counter that governs iteration termination").  
- Each explanation MUST be **self-contained** so it can be understood in isolation, without looking at [Target Code].  
4. Finally, provide the self-contained variable functionality explanations in valid JSON format with the following schema, including the leading and trailing "```json" and "```" 
```json
{
        "<variable_1_name>": string,  // the self-contained explanation of variable_1
        "<variable_2_name>": string,  // the self-contained explanation of variable_2
        ...
        "<variable_n_name>": string  // the self-contained explanation of variable_n
}
```
\end{lstlisting}

\begin{lstlisting}[caption={Prompt used for verification. Together with the one-shot example retrieved from the RAG DB (i.e., Verification CoT for \texttt{\{target\_cve\}}), this prompt verifies whether \texttt{[Target Code]} contains a vulnerability pattern similar to \texttt{\{target\_cve\}}.}]
**Role**: You are an expert software security engineer. Your goal is to analyze the user-provided [Target Code] to determine if it contains a vulnerability of type {target_cwe_type}, similar to {target_cve}. Focus on variables and functions with roles relevant to this vulnerability in [Target Code]. Perform the followings step by step and show the reasoning in each step. Start answering with "Let's think step-by-step."
---
**Task Overview**:
[Vulnerability-Related Variables]
{anonymized variables' description from rag-db for target_cve}
[Vulnerability-Related Functions]
{anonymized functions' description from rag-db target_cve}

Perform the followings step by step and show the reasoning in each step. Start answering with "Let's think step-by-step."
1) Using [Variable Mapping] and [Function Mapping], verify if {target_cwe_type} exists in [Target Code].
2) Based on the analysis in Step 1, identify the root cause of {target_cwe_type} within [Target Code].
3) Provide the results in the following schema, including the leading and trailing "```json" and "```" 
```json
{
        "result": boolean,  // the result of vulnerability verification (true = the vulnerability exists, false = the vulnerability does not exist)
        "cot" : string // the thinking process for the vulnerability verification (Step 1)
        "root_cause": string // the root cause of the vulnerability of {example_target_cwe_type} within [Target Code] (Step 2)
        
}
```
\end{lstlisting}

\begin{lstlisting}[caption={Prompt used for patching. Together with the one-shot example retrieved from the RAG DB (i.e., Patch CoT for \texttt{\{target\_cve\}}), this prompt patches \texttt{\{Target Code\}} to remove the vulnerability pattern similar to \texttt{\{target\_cve\}}.}]
**Role**: You are an expert software security engineer. Your goal is to patch the [Target Code] having a vulnerability of {target_cwe_type}, similar to {target_cve}. Focus on the given mappings of each symbolic variables and functions provided by user with [Variable Mapping] and [Function Mapping].
---
**Task Overview**:
[Vulnerability-Related Variables]
{anonymized variables' description from rag-db for target_cve}
[Vulnerability-Related Functions]
{anonymized functions' description from rag-db target_cve}

Perform the followings step by step and show the reasoning in each step. Start answering with "Let's think step-by-step."
1. Based on [Variable Mapping] and [Function Mapping], describe how to patch the [Target Code] for fixing {target_cwe_type} similar to {target_cve}.
2. Use the patch description from Step 1 to generate a patched code.
3. Provide the results in the following schema, including the leading and trailing "```json" and "```" 
```json
{
        "cot" : string // the thinking process for the vulnerability patching (Step 1)
        "patched_code": string // the patched code (Step 2)
        
}
```
\end{lstlisting}

\subsection{Acutal Prompts for CVE-2025-21671}

\label{sec:appendix_prompt}
\begin{figure*}[t]
    \centering
    
    % \hspace*{-0.11\linewidth}
    \includegraphics[width=0.796\linewidth]{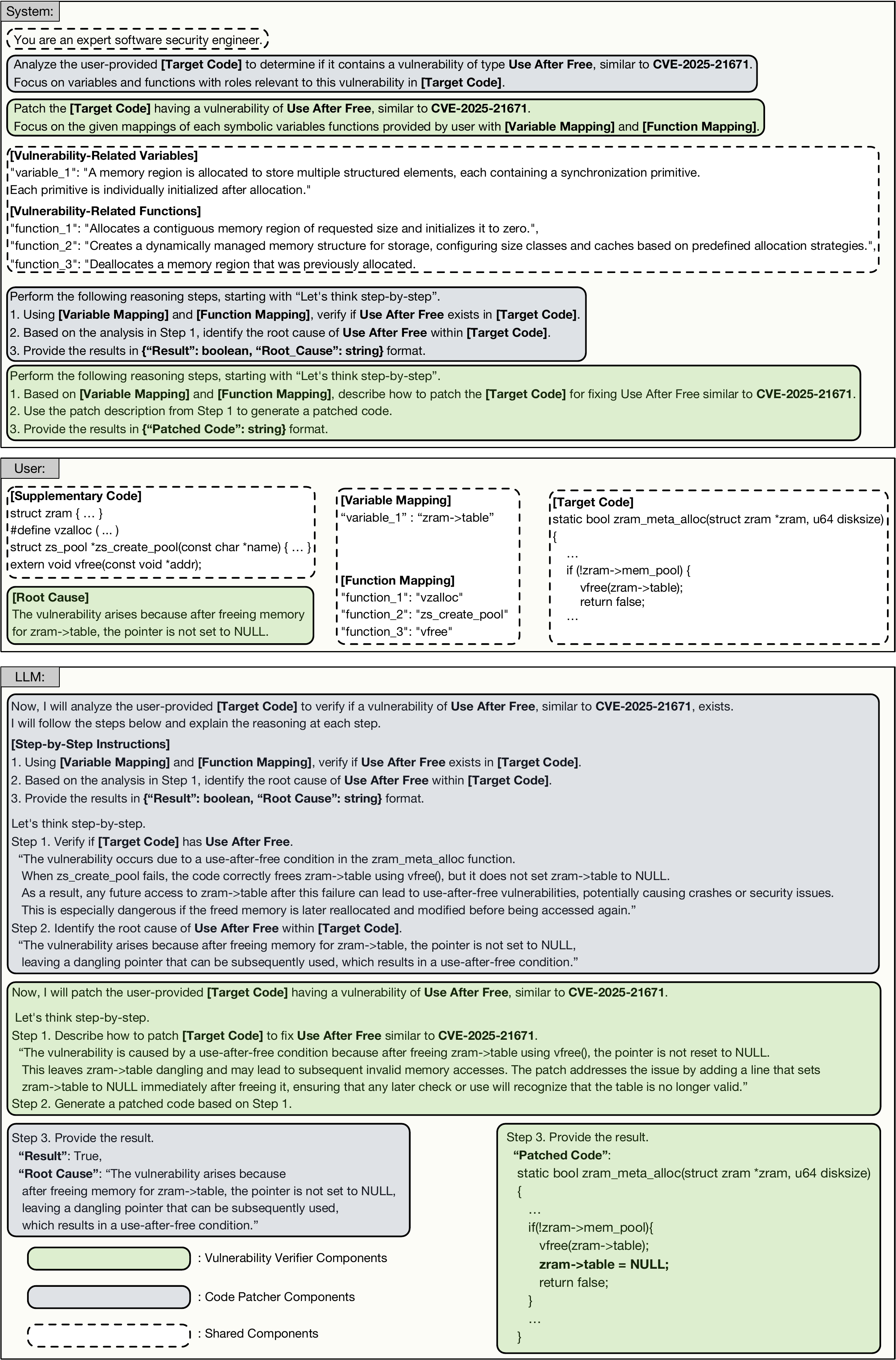}
    \caption{Example verification and patch prompt for CVE-2025-21671.}
    \label{fig:verification_and_patch_prompt}
\end{figure*}

\end{document}